\documentclass[fleqn,12pt,letterpaper]{article}

\usepackage{amsfonts,amssymb,amsmath,amsthm,color,float,graphicx,natbib,rotating}
\usepackage{epstopdf}
\usepackage[margin=2.54cm]{geometry}
\definecolor{darkred}{rgb}{0.5,0.2,0.2}
\usepackage[allcolors=darkred,bookmarks=false,colorlinks=true]{hyperref}
\usepackage{booktabs}
\usepackage{rotating} 
\usepackage{pdflscape}
\theoremstyle{plain}

\newtheorem{definition}{Definition}
\theoremstyle{definition}

\newtheorem{remark}{Remark}
\usepackage{enumitem}
 \usepackage{multirow}

  \usepackage{float}
\newcommand{\E}{\mathbb{E}}

\numberwithin{equation}{section}
\makeatletter
\newcommand*\rel@kern[1]{\kern#1\dimexpr\macc@kerna}
\newcommand*\widebar[1]{%
  \begingroup
  \def\mathaccent##1##2{%
    \rel@kern{0.8}%
    \overline{\rel@kern{-0.8}\macc@nucleus\rel@kern{0.2}}%
    \rel@kern{-0.2}%
  }%
  \macc@depth\@ne
  \let\math@bgroup\@empty \let\math@egroup\macc@set@skewchar
  \mathsurround\z@ \frozen@everymath{\mathgroup\macc@group\relax}%
  \macc@set@skewchar\relax
  \let\mathaccentV\macc@nested@a
  \macc@nested@a\relax111{#1}%
  \endgroup
}
\makeatother

\newlength{\sigwd}
\AtBeginDocument{\settowidth{\sigwd}{$^{***}$}}
\newcommand{\sig}[1]{\makebox[\sigwd][l]{$^{#1}$}}

\begin{document}

\title{The ``Rough'' HAR model\footnote{We are grateful to Peter R. Hansen, Thomas K. Kloster, Roberto Renò, Shuping Shi and Ke Zhu for their helpful discussions and comments. We also thank the participants of the Financial Econometrics Workshop at the University of Macau, the 2026 Vienna-Copenhagen Conference on Financial Econometrics, and the 2026 annual SoFiE Conference for their valuable feedback. MB acknowledges support from the Aarhus Center for Econometrics (ACE), funded by the Danish National Research Foundation, grant number DNRF186.}}

\author{Mikkel Bennedsen\thanks{Aarhus University, Department of Economics and Business Economics, Denmark.} $^,$\thanks{Center for Research in Energy: Economics and Markets (CoRE), Aarhus University, Denmark.} $^,$\thanks{Aarhus Center for Econometrics (ACE),  Aarhus University, Denmark.}
\and Kim Christensen\footnotemark[2] $^,$\footnotemark[3] $^,$\thanks{Research fellow at the Danish Finance Institute (DFI).}
\and Peter Korsbakke Christensen\footnotemark[2] $^,$\footnotemark[3] \and  Jun Yu\thanks{University of Macau, Faculty of Business Administration, Macau.} \and  Chen Zhang\thanks{Sun Yat-sen University, Lingnan College, China.} }

\date{September 2026}

\maketitle

\vspace*{-1.20cm}

\begin{abstract}
This paper proposes discrete-time approximations to rough continuous-time models of realized variance (RV). The leading rough models can be viewed as autoregressive processes driven by fractional Gaussian noise. We show that the Wold representation of this noise concentrates its dependence at the first lag when the Hurst parameter is below one half. Augmenting the autoregressive (AR) and heterogeneous autoregressive (HAR) models with a first-order moving-average (MA(1)) component therefore approximates the roughness, and the MA coefficient maps almost linearly into the Hurst parameter. We refer to these extensions as the ``rough'' AR and ``rough'' HAR models. Estimating them on the log RV of ten ETFs, we find negative MA coefficients for every asset, and the implied Hurst parameters align closely with the estimates from the continuous-time models. In the HAR literature, the negative MA(1) component is a significant feature that has been largely overlooked. In out-of-sample comparisons, the ``rough'' models outperform their classical counterparts for nearly every asset and horizon, with the largest gains at short horizons, and their accuracy is comparable to that of the rough continuous-time models but much easier to estimate by standard off-the-shelf software.

\bigskip \noindent \textbf{JEL Classification}: C10; C58; C80.

\medskip \noindent \textbf{Keywords}: Continuous-time models, roughness, HAR model, realized variance. 

\end{abstract}

\vfill

\thispagestyle{empty}

\pagebreak

\section{Introduction}

Volatility is a central input to asset pricing, portfolio allocation, and risk management. Since the seminal contributions of \citet{engle:82a}, \citet{bollerslev:86a}, and \citet{taylor:82a}, a vast literature has developed GARCH and stochastic volatility models to describe its dynamics (\citealp{bauwen:06,asai:06}).

The advent of high-frequency data has changed how volatility is measured. Realized variance (RV), the sum of squared intraday returns, has emerged as the standard nonparametric estimator of daily variance \citep{andersen-bollerslev-diebold-labys:01a,barndorff-nielsen-shephard:02a}. It has long been known that RV is highly persistent across a wide range of financial assets, where estimates indicate that the series possesses formal long memory, in the sense that the autocorrelation function is non-integrable \citep{andersen-bollerslev-diebold-labys:03a}. \citet{andersen-bollerslev-diebold-labys:03a} suggest capturing this persistence with autoregressive fractionally integrated moving average (ARFIMA) models \citep{granger:80a,granger-joyeux:80a}. In practice, however, the ARFIMA framework is nontrivial to estimate and extend \citep{corsi:09a}. Instead, \citet{corsi:09a} proposes the heterogeneous autoregressive (HAR) model as a simple alternative, in which an additive cascade of volatility components corresponds to trader types with different horizons. Although not formally a long-memory model, the HAR model reproduces the strong persistence of RV. It has become the benchmark for volatility forecasting, and numerous extensions have been proposed (e.g., \citealp{corsi-reno:12a,patton-sheppard:15a,bollerslev-patton-quaedvlieg:16a,bollerslev-li-patton-quaedvlieg:20a}).

A more recent strand of the literature, pioneered by \citet{gatheral-jaisson-rosenbaum:18a}, emphasizes the local behavior of volatility. In this rough volatility literature, the estimated roughness parameter of volatility is typically well below $0$.  The sample paths of volatility are then rougher than those of a standard Brownian motion. This finding is stable across stock indices, ETFs, and individual stocks, and it extends to spot volatility and trading volume \citep{fukasawa-takabatake-westphal:22a,bolko-christensen-pakkanen-veliyev:23a,wang-xiao-yu:23a,shi-yu-zhang:24a,chong-todorov:25a,takabatake-yu-zhang:25a,bennedsen-christensen-christensen:26a}. Several studies seek the economic origin of roughness in the microstructure of trading (e.g., \citealp{euch-fukasawa-rosenbaum:18a,jusselin-rosenbaum:20a}), and rough volatility models are widely used in derivative pricing and hedging \citep{bayer-friz-gatheral:16a,euch-rosenbaum:18a}. Several papers have demonstrated that continuous-time rough models perform very well for forecasting realized variance, even outperforming the benchmark HAR models \citep{bennedsen-lunde-pakkanen:22a,wang-xiao-yu:23a}.

The continuous-time formulation comes at a cost, however. The leading rough
models, such as fBm and the fractional Ornstein--Uhlenbeck (fOU) process, are
non-Markovian, and not even semimartingales. Their estimation is therefore
demanding, and efficient methods exist only for a limited set of processes
\citep{wang-xiao-yu:23a,bennedsen-christensen-christensen:26a,wang-xiao-yu-zhang:25a,takabatake-yu-zhang:25a}.
Extensions beyond the univariate model must be engineered case by case, and
the multivariate problem remains largely open
\citep{dugo-giorgio-Pigato:2025a,bibinger-yu-zhang:2025a}. The discrete-time
framework we propose instead inherits its extensions off the shelf from the
ARMA toolbox. We elaborate on these issues in Section
\ref{subsec:limitations}. The situation is reminiscent of the position of
ARFIMA models in the long-memory literature, which motivated the HAR model in
the first place.

In this paper, we propose a simple discrete-time approximation to the rough continuous-time models. The starting point is that fBm and the fOU process can be viewed as an autoregressive process of order one (AR(1)) driven by fractional Gaussian noise (fGn). The task is then to approximate the fGn. We show that the Wold representation of fGn concentrates its dependence at the first lag when the Hurst parameter is below $0.5$. The first Wold coefficient is negative, and close to linear in the Hurst parameter. Hence, a first-order moving average (MA(1)) with a negative coefficient is a natural proxy for fGn. Combining the MA(1) component with an AR(1) or a HAR structure yields what we call the ``rough'' AR and ``rough'' HAR models. The models are linear and parsimonious, and replicate both the persistence and the roughness of volatility. They can be estimated by maximum likelihood, with the likelihood evaluated using the Kalman filter. They also extend as easily as any ARMA-type model, for example, with additional predictors or time-varying parameters. Finally, the MA term is easy to interpret. When its coefficient is negative, consecutive movements in volatility tend to partially reverse. In Section \ref{sec:motivation}, we relate this behavior to a microstructural explanation of roughness, in which large orders are split up and executed gradually over time.

A simulation study supports the approximation. When we fit the ``rough'' models to data simulated from fBm and fOU, the estimated MA coefficient recovers the theoretical mapping to the Hurst parameter.  We then apply the models to the log RV of ten ETFs. The estimated MA(1) coefficients are negative for every asset. Through the mapping between the MA coefficient and the Hurst parameter, the estimates imply Hurst parameters well below $0.5$. This agrees with the evidence from continuous-time fractional models. However, in the HAR literature, the negative MA(1) component appears to have been largely overlooked. In out-of-sample comparisons, the ``rough'' models outperform their classical counterparts for nearly every asset and horizon, with substantial gains at short horizons, under both the mean squared error (MSE) and QLIKE loss functions. Their accuracy is comparable to that of more complex continuous-time rough models, the fBm and fOU.  As a robustness check, we repeat the forecasting comparison for forty individual stocks, with similar conclusions. 

 A negative MA coefficient in RV can also arise from a different mechanism, namely measurement error. RV is a noisy estimate of the latent integrated variance, and the noise induces an ARMA structure in the observed series \citep{barndorff-nielsen-shephard:02a,meddahi:03a,andersen-bollerslev-meddahi:04a}. If the latent log variance follows an AR(1) process and the noise is serially independent, the observed series follows an ARMA(1,1) with a negative MA coefficient.  \citet{hansen-lunde:14a} exploit this structure and estimate the persistence of the latent process with instrumental variables. The reduced form of our ``rough'' AR model is observationally similar to these specifications at first glance; however, its underlying interpretation is fundamentally different. If measurement noise is the only, or the primary, contributor to the observed roughness, we would expect two things. First, our rough models should perform similarly to the HARQ model of \citet{bollerslev-patton-quaedvlieg:16a}, which is specifically designed to account for measurement error. Second, the state-space formulation of \citet{asai-mcaleer-medeiros:12a} suggests that, after accounting for measurement error, the latent log variance should be smooth rather than rough. Our results provide strong evidence against both expectations. The HARQ model fails to deliver the forecasting gains achieved by our rough models, while the estimated latent log variance remains rough even after accounting for measurement error.

 



The remainder of the paper is organized as follows. Section \ref{sec:motivation} motivates the new class of models. Section \ref{sec:simulation} studies the quality of the approximation in simulations. Section \ref{sec:empirical} contains the empirical analysis. Section \ref{sec:conclusion} concludes. The appendices collect additional empirical and technical details.

\section{Simple Approximate Model for Realized Variance Forecasting}
\label{sec:motivation}

\subsection{Continuous-time fractional models}
\label{subsec:ctmodels}

Since the seminal work of \citet{gatheral-jaisson-rosenbaum:18a}, considerable effort has been devoted to modeling volatility with continuous-time fractional processes. In this subsection, we present the two representative models that we work with throughout the paper. We organize the presentation around two properties of a stochastic process. The first property concerns its behavior over short time scales.


\begin{definition}[Roughness]
\label{def:rough}
A stochastic process $\{y_t : t \in \mathbb{R}\}$ with stationary increments has roughness index $\alpha \in (-1/2, 1/2)$ if its variogram satisfies
\begin{equation*}
\E\left[ \left( y_{t+h} - y_t \right)^2 \right] \sim L(h) |h|^{2\alpha+1}, \qquad |h| \to 0,
\end{equation*}
where $L$ is a function that is slowly varying at zero.
\end{definition}

We call a process \emph{rough} if $\alpha < 0$ and \emph{smooth} if $\alpha > 0$. A Brownian motion has $\alpha = 0$. For a Gaussian process, the Kolmogorov--Chentsov theorem links the variogram to the regularity of the sample paths. A process with roughness index $\alpha$ admits a modification whose paths are Hölder continuous of any order $\gamma < \alpha + 1/2$. A rough process therefore has more erratic sample paths than a Brownian motion, and a smooth process has more regular ones. The second property concerns the behavior over long time scales.

\begin{definition}[Long memory]
\label{def:lm}
A stationary stochastic process $\{y_t: t\in\mathbb{R}\}$ has long memory of degree $\beta \in (0, 1]$ if its autocorrelation function $\rho_h = Corr(y_{t+h},y_t)$ satisfies
\begin{equation*}
\rho_h \sim L_{\infty}(h) |h|^{-\beta}, \qquad |h| \to \infty,
\end{equation*}
where $L_{\infty}$ is a function that is slowly varying at infinity.
\end{definition}

The restriction on $\beta$ implies that the autocorrelation function is not integrable. If the autocorrelation function is integrable, we say that the process has short memory. Roughness is a local property and long memory is a global one. A priori, the two are unrelated. 

The first model we consider is fBm, which goes back to \citet{mandelbrot-vanness:68} and which \citet{gatheral-jaisson-rosenbaum:18a} introduced into volatility modeling. Let $\{y_t : t\in\mathbb{R}\}$ denote a scaled fBm:
\begin{equation*}
y_t=\sigma B^{H}_t,
\end{equation*}
where $\sigma>0$ is a scale parameter, $H\in(0,1)$ is the Hurst parameter, and $B^{H}_t$ admits the following moving-average representation in terms of a standard Brownian motion $B_t$:
\begin{equation}
B^{H}_t=\frac{1}{\Gamma (H+0.5)}\left\{ \int_{-\infty }^{0}\left[ \left(
t-s\right) ^{H-0.5}-(-s)^{H-0.5}\right] d B_s+\int_{0}^{t}\left(
t-s\right) ^{H-0.5} d B_s\right\}.
\label{def-fbm}
\end{equation}%
When $H=0.5$, $B^{H}_t=B_t$ is a standard Brownian motion. The variogram of the scaled fBm is proportional to $|h|^{2H}$, so Definition \ref{def:rough} applies with $\alpha = H - 1/2$. The process is rough exactly when $H < 1/2$. For our purposes, a second representation is more instructive. Sampled at integer times, fBm admits a unit-root structure,
\begin{align*}
    B^{H}_t=B^{H}_{t-1}+x_t,
\end{align*}
where $x_t = B^{H}_t-B^{H}_{t-1}$ is the fGn. The fGn is stationary and its autocovariance function is given by
\begin{align}
Cov(x_{t+h},x_t) = \frac{1}{2}\left(|h+1|^{2H}-2|h|^{2H}+|h-1|^{2H}\right),\qquad h\in\mathbb{Z}.
 \label{acf_fgn}
\end{align}
When $H<1/2$, the increments $x_t$ are negatively correlated. This produces the frequent reversals that make the sample path rough. At long lags, the autocovariance in \eqref{acf_fgn} decays at the rate $|h|^{2H-2}$, so the increments have long memory in the sense of Definition \ref{def:lm} if and only if $H > 1/2$. In the fBm, therefore, a single parameter controls both ends of the autocorrelation, and roughness and long memory are coupled.

The fBm is nonstationary and cannot capture the mean reversion of volatility. To address this limitation, \citet{wang-xiao-yu:23a} propose modeling RV with the fOU process, defined as the solution to
\begin{equation*}
dy_{t} = \kappa \left( \mu - y_{t} \right) dt + \sigma \, dB^{H}_t,
\end{equation*}
where $\kappa>0$, $\sigma > 0$, $\mu \in \mathbb{R}$, and $B^{H}_t$ is the fBm in \eqref{def-fbm}. The fOU process has a unique path-wise solution, given by
\begin{equation*}
y_{t}=e^{-\kappa t}y_{0}+\left( 1-e^{-\kappa t}\right) \mu
+\int_{0}^{t}\sigma e^{-\kappa (t-s)}dB^{H}_s.
\end{equation*}
The exact discrete-time model of the fOU process, evaluated at time points $(i-1)\Delta$ and $i\Delta$, is given by:
\begin{equation*}
y_{i\Delta} = e^{-\kappa\Delta} y_{(i-1)\Delta} + \left(1 - e^{-\kappa\Delta}\right) \mu + \eta_{i\Delta}
\quad \text{with} \quad
\eta_{i\Delta} = \sigma \int_{(i-1)\Delta}^{i\Delta} e^{-\kappa(i\Delta - s)} \, dB^H_s,
\end{equation*}
where $\Delta$ is the sampling interval. The fOU process reduces to the classical OU process when $H=0.5$ and to fBm when $\kappa=0$. In contrast to fBm, it is stationary when $\kappa > 0$ under conditions on the initial value. Mean reversion does not, however, resolve the coupling of the time scales. Over a short interval of length $h$, the mean reversion contributes a term of order $h$ to the increment, while the driving fBm contributes a term of order $|h|^{H}$. The latter dominates near the origin, so the variogram of the fOU process behaves like $\sigma^2 |h|^{2H}$, and Definition \ref{def:rough} applies with $\alpha = H - 1/2$, exactly as for fBm. At the long end, \citet{cheridito-kawaguchi-maejima:03a} show that for any $\kappa > 0$, the autocovariances of the fOU process behave like those of the driving fGn, so the correlation decays at the rate $|h|^{2H-2}$. The Hurst parameter thus still governs both time scales. Like fBm, the fOU process can be rough or have long memory, but not both. When $\kappa$ is positive but small and $H < 1/2$, it nevertheless mimics both features in finite samples, which is the configuration relevant for volatility. The two properties can also be decoupled altogether. \citet{bennedsen-lunde-pakkanen:22a} and \citet{bennedsen-christensen-christensen:26a} model volatility with the Cauchy class of \citet{gneiting-schlather:04a}, in which roughness and memory are governed by separate parameters, and they document both features in the data.

Why is volatility rough in the first place? The leading explanation operates through the microstructure of trading. Large orders are rarely executed at once. Instead, they are split into many small orders and fed to the market gradually, which makes the signs of trades positively autocorrelated over long stretches of time \citep{lillo-farmer:04a,toth-palit-lillo-farmer:15a}. At the same time, most order flow is a reaction to other order flow, so markets operate in a regime of high endogeneity \citep{hardiman-bercot-bouchaud:13a}. \citet{jaisson-rosenbaum:16a} model the arrival of orders as a nearly unstable Hawkes process \citep{hawkes:71a} with a heavy-tailed kernel, and they show that the induced volatility converges to a rough process with a small Hurst parameter. \citet{euch-fukasawa-rosenbaum:18a} build the microstructural foundations of the rough Heston model on the same mechanism, and \citet{jusselin-rosenbaum:20a} show that the absence of arbitrage ties power-law market impact to rough volatility. This suggests that roughness is not merely a statistical artifact, but is a real feature of volatility, which derives from the order splitting under high endogeneity. We stress that these results concern the continuous-time limit of the trading process. They explain why the fBm and fOU models of this section are specified as rough when modeling volatility. 

\subsection{Limitations in empirical work}
\label{subsec:limitations}

The models above are natural benchmarks for volatility. They are, however, difficult to apply in empirical work. There are two barriers, estimation and extension. We review them in turn, since they motivate the approximation that follows. 

The first challenge is estimation. For the fOU process, \citet{wang-xiao-yu:23a} propose a two-stage procedure based on moments, and \citet{bennedsen-christensen-christensen:26a} introduce a maximum composite likelihood (MCL) approach. More recently, \citet{shi-yu-zhang:24a} develop a Whittle-type likelihood method, and \citet{wang-xiao-yu-zhang:25a} develop exact maximum likelihood. For more general continuous-time fractional models, efficient techniques remain scarce. The most general estimation frameworks currently available are those of \citet{bennedsen-christensen-christensen:26a} and \citet{takabatake-yu-zhang:25a}. Even there, several challenges remain, including the derivation of closed-form spectral densities and the verification of regularity conditions for specific processes. The full likelihood also carries a heavy computational burden, because the covariance matrix of a non-Markovian Gaussian process must be factorized at every evaluation.\footnote{Composite likelihood methods reduce this cost substantially, at the price of some efficiency \citep{bennedsen-christensen-christensen:26a}.}

The second barrier is extension. The continuous-time fractional models are formulated in terms of their own history alone. The forecasting literature, by contrast, exploits additional predictors, such as realized semivariance and realized semicovariance. Such predictors can be appended linearly to a Gaussian fractional model, and the resulting regression with fractional errors can be estimated by maximum likelihood. Beyond this linear-in-mean case, however, each extension must be engineered from scratch. A structural break or deterministic time variation in the parameters destroys stationarity. The spectral density and the Toeplitz covariance structure are then lost, and the Gaussian likelihood must be built and factorized case by case. Regime switching and stochastic parameter variation are harder still. The likelihood then averages over the paths of a latent state, and because the processes are non-Markovian, there is no finite-dimensional filter to compute this average recursively. 

Multivariate extensions are especially challenging. Modeling the volatilities of several assets at once requires a vector of fractional processes with a valid joint law. \citet{bibinger-yu-zhang:2025a} and \citet{dugo-giorgio-Pigato:2025a} have recently proposed such multivariate models, and their specifications require restrictions to guarantee non-degeneracy. On the estimation side, \citet{bibinger-yu-zhang:2025a} develop a method-of-moments estimator based on first-order increments, and \citet{dugo-giorgio-Pigato:2025a} propose a feasible, albeit inefficient, generalized method of moments estimator. Likelihood-based efficient estimation remains unavailable. 

 None of these steps is impossible. Each one, however, gives up part of what makes the continuous-time models attractive, and none comes with the off-the-shelf toolbox. As noted by \citet{corsi:09a}, ``fractionally integrated models are
nontrivial to estimate and not easily extendible to multivariate processes.'' What that literature needed was the HAR model. In the context of rough volatility, we propose a similar approximate model that facilitates estimation and extension.

\subsection{Relation between rough volatility models and MA specifications}
\label{subsec:arma_motivation}

 Revisiting classical continuous-time fractional models, such as fBm and the fOU process, reveals that they can be interpreted as an AR(1) process driven by an (approximate) fGn. Consequently, constructing approximate models for rough volatility models reduces to approximating the driving fGn. Motivated by the autocorrelation structure of fGn, in which the first-order autocorrelation is substantially negative while higher-order autocorrelations decay rapidly, we consider an MA-type approximation. This approximation is further justified theoretically.

 


We begin with the discrete-time representation of the fOU process, which nests fBm as a special case. Without loss of generality, we consider a centered fOU process (with $\mu=0$  known), satisfying
\begin{equation}
y_{i\Delta} = e^{-\kappa\Delta} y_{(i-1)\Delta}  + \eta_{i\Delta}
\quad \text{with} \quad
\eta_{i\Delta} = \sigma \int_{(i-1)\Delta}^{i\Delta} e^{-\kappa(i\Delta - s)} \, dB^H_s.
 \label{eq:fou_discrete}
\end{equation}
For $\kappa = 0$, the kernel is constant, and $\eta_{i\Delta}$ reduces to an increment of $B^{H}$, which is the fBm case. For $\kappa > 0$, $\eta_{i\Delta}$ is instead a filtered version of the increments of $B^{H}$. The filtering is mild over a short sampling interval. The kernel deviates from one by a term of order $\Delta$, and integrating this deviation against $B^{H}$ over an interval of length $\Delta$ produces a term of order $\Delta^{1+H}$. Hence,
\begin{equation*}
\eta_{i\Delta}
= \sigma \left( B_{i\Delta}^{H} - B_{(i-1)\Delta}^{H} \right)
+ O_p\!\left( \Delta^{1+H} \right);
\end{equation*}
see also \citet{wang-xiao-yu:23a}. The leading term is the increment of fBm over the sampling interval, which is a scaled fGn of order $\Delta^{H}$. The approximation error is therefore smaller than the noise itself by a factor of $\Delta$. Approximating the fBm or fOU processes thus reduces to approximating the fGn.

The dependence structure of fGn is distinctive. Rough processes are characterized by a sharp drop in the autocovariance function around the origin. From the autocovariance function in \eqref{acf_fgn}, the first-order autocorrelation equals $2^{2H-1} - 1$, or about $-0.34$ at $H = 0.2$, which is the center of the empirically relevant range. The autocorrelations at higher lags are negative as well, but they decay hyperbolically and are an order of magnitude smaller. A sharp drop after the first lag is also the signature of a low-order MA process. A low-order MA is therefore a natural candidate for approximating the driving fGn. The remainder of this subsection makes the approximation precise.

Write $X_t$ for the fGn. Its spectral measure is absolutely continuous, and
its spectral density $f$ satisfies Szegö's condition
$\log f \in L^1([-\pi,\pi])$. Together, these two properties imply that
$X_t$ is purely nondeterministic, so it admits the Wold decomposition
\begin{equation}
    X_t \;=\; \sum_{j=0}^{\infty} c_j\,\varepsilon_{t-j},
    \qquad c_0=1,
    \qquad \sum_{j=0}^{\infty} c_j^2 < \infty,
    \label{eq:wold}
\end{equation}
where $\{\varepsilon_t\}$ is a white-noise innovation sequence with variance
$\sigma_\varepsilon^2$. Truncating the expansion after lag $q$ gives
$X_t^{(q)} := \sum_{j=0}^{q} c_j\,\varepsilon_{t-j}$. Because the
innovations are orthogonal, $X_t^{(q)}$ is the projection of $X_t$ onto the
closed linear span of $\{\varepsilon_t, \ldots, \varepsilon_{t-q}\}$. It is
therefore the $L^2$-optimal approximation to $X_t$ among linear combinations
of the current and $q$ lagged innovations, and its mean squared error is
\begin{equation*}
    \mathbb{E}\!\left[(X_t - X_t^{(q)})^2\right]
    \;=\;
    \sigma_\varepsilon^2\sum_{j=q+1}^{\infty} c_j^2.
\end{equation*}
This optimality holds at every $q$, so it does not by itself justify a short
truncation. What matters is how fast the coefficients decay.

For fGn, the Wold coefficients decay at a hyperbolic rate. It
can be shown that
\begin{equation}
    c_j \;\sim\; K(H)\, j^{\,H-\frac{3}{2}}
    \qquad\text{as } j\to\infty,
    \label{eq:ck_asympt}
\end{equation}
for a constant $K(H)\neq 0$, see Theorem~6 of \citet{mcleod:98a}. When
$H\in(0,1/2)$, the exponent satisfies $H-\tfrac{3}{2}<-1$, implying that
$\{c_j\}$ is absolutely summable and the tail $\sum_{j>q} c_j^2$ decays
quickly. This supports a small $q$ in the rough-volatility range.

To quantify the approximation, we compute the Wold coefficients numerically from the spectral density of fGn. \citet{shi-yu-zhang:25b} show that this spectral density admits a closed-form representation in terms of the Hurwitz zeta function,
\begin{equation*}
f(\lambda)
= 2C_H\bigl(1-\cos(\lambda)\bigr)(2\pi)^{-1-2H}
\left[
\zeta\!\left(1+2H,\,1-\frac{\lambda}{2\pi}\right)
+\zeta\!\left(1+2H,\,\frac{\lambda}{2\pi}\right)
\right],
\end{equation*}
where \(C_H=\sigma^{2}(2\pi)^{-1}\Gamma(2H+1)\sin(\pi H)\) and
\(\zeta(s,q)=\sum_{j=0}^{\infty}(j+q)^{-s}\) is the Hurwitz zeta function.

Under Szegö's condition (see, e.g., \citealp{bingham:2012a}), $f$ admits a Szegö
spectral factorization. That is, there exists an outer function $C$  that is analytic for $|z|<1$ with
$C(0)=1$ such that
\begin{equation*}
    f(\lambda)
    \;=\;
    \frac{\sigma_\varepsilon^{2}}{2\pi}\,\bigl|C(e^{-i\lambda})\bigr|^{2},
    \qquad \lambda\in[-\pi,\pi]\ \text{(a.e.)},
\end{equation*}
where $\sigma_\varepsilon^{2}$ is the innovation variance. Moreover, $C$ admits the power-series
representation $C(z)=\sum_{j\ge 0} c_j z^j$, and the coefficients $\{c_j\}$ coincide with the innovations
filter in the Wold representation \eqref{eq:wold}. The link between $C$
and the  cepstral coefficients\footnote{The cepstral coefficients are the Fourier coefficients of the log-spectral density.} $a_k$ of $\log f$ is summarized in Appendix~\ref{app:szego_factor}. We compute the Fourier coefficients
\begin{equation*}
    a_k
    \;=\;
    \frac{1}{2\pi}\int_{-\pi}^{\pi} \log f(\lambda)\,e^{-ik\lambda}\,d\lambda,
\end{equation*}
and then obtain $\{c_j\}$ from the standard cepstral recursion (see, e.g.,
\citealp{pourahmadi:01a,bingham:2012a})
\begin{equation}
    c_{k+1}
    \;=\;
    \sum_{j=0}^{k}\left(1-\frac{j}{k+1}\right)a_{k+1-j}\,c_j,
    \qquad k=0,1,2,\ldots,
    \label{eq:pourahmadi_recursion}
\end{equation}
with $c_0=1$. The resulting coefficients are plotted in Figure \ref{fig:H_c_figure} and tabulated in Appendix \ref{app:szego_factor}.

Figure \ref{fig:H_c_figure} summarizes the behavior of the coefficients in the rough range. Panel A plots the first five MA coefficients over $H \in [0.05, 0.5]$. The first coefficient $c_1$ is negative, consistent with the negative short-run dependence implied by \eqref{acf_fgn} when $H<1/2$. It also dominates the higher-order coefficients, whose rapid decay reflects the asymptotic behavior in \eqref{eq:ck_asympt}. The relationship between $c_1$ and the Hurst parameter is monotonic and almost linear, which is both interesting and surprising. This relationship suggests that $c_1$ is particularly informative about the Hurst parameter. As $H$ varies from $0$ to $0.5$, the coefficient ranges approximately from $-1$ to $0$. In contrast, such a relationship is not observed between $d$ in the ARFIMA model and the Hurst parameter $H$ in \citep{shi-yu-zhang:25a}, although the two parameters are asymptotically related by $d=H-\frac{1}{2}$ \citep{tanaka:2013,Wang-Yu}. In particular, when $H$ falls below a certain threshold, say $H=0.2$, the estimated $d$ tends to hit the lower boundary of $-0.5$, rather than varying smoothly around $-0.3$ as suggested by the asymptotic relationship.  Panel B plots the ratio of $c_1^2$ to $\sum_{j=1}^\infty c_j^2$, which measures the share of the variance contributed by the lagged terms of the MA(\(\infty\)) representation that is explained by the first lag.\footnote{In practice, we are of course unable to compute the exact value of the infinite series, so we approximate it using the first 100 terms. Given the rapid decay of the coefficients, the approximation error is negligible.} This share is large, especially in the empirically relevant rough range.

The first lag therefore does almost all of the work. Together with the
optimality of the truncation, this justifies cutting the expansion at the
first lag. We set $q = 1$ \footnote{We do not argue that the best approximation must necessarily be an MA(1) model. Rather, our point is that this parsimonious specification is highly informative about the Hurst parameter and can explain a substantial fraction of the variation in the data. Higher-order specifications, such as MA(2) and MA(3), can also be readily implemented. Interestingly, when we experiment with the data, we find that the MA(1) specification achieves the smallest BIC among the MA(1), MA(2), and MA(3) specifications for the log-RV of SPY.} and approximate the fGn by
\begin{equation*}
    X_t \;\approx\; \varepsilon_t + c_1\,\varepsilon_{t-1}.
\end{equation*}

\begin{figure}[H]
\begin{center}
\caption{Relation between $c$ and $H$}
\label{fig:H_c_figure}
\begin{tabular}{cc}
\small{Panel A: $c-H$ relation.} & \small{Panel B: Share of variance.} \\
\includegraphics[width=0.48\textwidth]{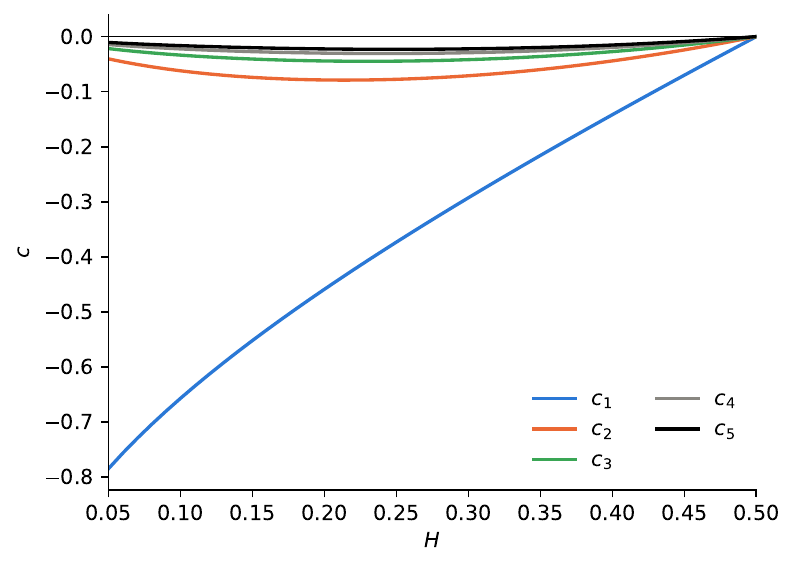} &
\includegraphics[width=0.48\textwidth]{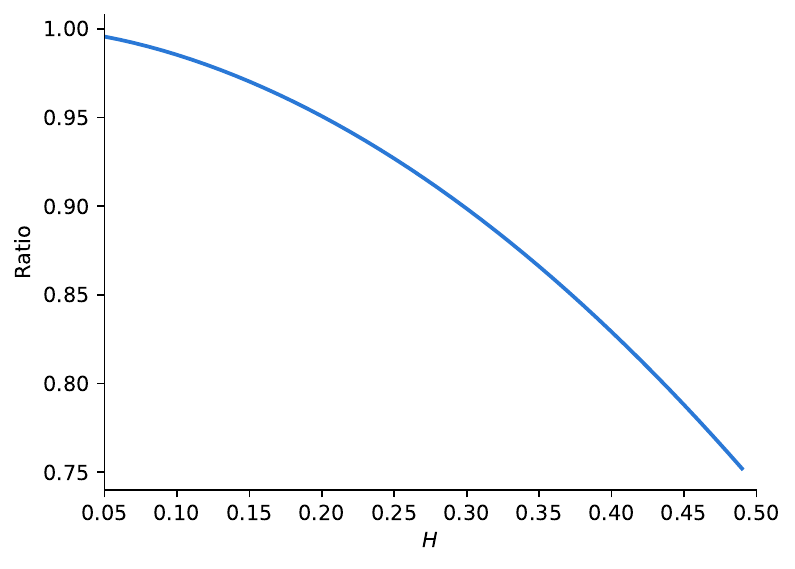} \\
\end{tabular}
\begin{scriptsize}
\parbox{\textwidth}{\emph{Note.} We compute the MA coefficients $c_j$ for $j=0,1,\ldots,100$ for the fractional Gaussian noise process for different $H$ values in the range $[0.05, 0.5]$. In Panel A, we show the first five MA coefficients over the range of $H$ values, while Panel B shows the squared first MA coefficient, $c_1^2$, divided by the sum of squared MA coefficients, $\sum_{j=1}^{100}c_j^2$.}
\end{scriptsize}
\end{center}
\end{figure}

\subsection{``Rough'' models}

As shown above, the MA(1) process has the potential to mimic the behavior of fGn. Since fBm and the fOU process can be approximately represented as autoregressive processes of order one driven by an fGn component, a natural approximation to the continuous-time fractional models emerges. We keep the autoregressive structure and replace the fGn by an MA(1) component. This approach is analogous to the work of \citet{corsi:09a}, where the HAR model is viewed as a discrete-time approximation to the ARFIMA model. This yields an ARMA(1,1) approximation:

\begin{align}
y_j = \omega + \rho y_{j-1} + \epsilon_j, \label{ar1-spec}
\end{align}
where $y_j$ denotes the observed log realized variance, $\omega$ is a constant, $\rho$ controls persistence, and $\epsilon_j = \varepsilon_j + \theta \varepsilon_{j-1}$ captures the roughness. $\varepsilon_j$ are i.i.d. normally distributed as $N(0,\sigma^2)$. When $\theta$ is negative, the model generates roughness and may be referred to as the \emph{``rough'' AR model}.\footnote{The name should not be taken literally. Roughness in the sense of Definition \ref{def:rough} concerns the variogram at infinitesimal lags, which is not defined for a process in discrete time. The model instead mimics the sharp drop of the autocovariance function at the first lag that rough processes display when sampled at the daily frequency.}

For the ``rough'' AR model in \eqref{ar1-spec}, the autocovariance function has a particularly simple form. The first autocovariance is
\begin{align}
    \gamma_1 = \sigma^2 \left[ (\theta + \rho) + \frac{(\theta + \rho)^2 \rho}{1 - \rho^2} \right],
    \label{acf_arma}
\end{align}
and the higher-order autocovariances decay geometrically, with $\gamma_k = \rho^{k-1} \gamma_1$ for $k > 1$.

The sign of $\theta$ is important. Roughness is a sample-path property of a stochastic process and manifests itself as a rapid decline in the autocorrelation function (ACF) over short time scales. Empirically, the autoregressive coefficient of log RV is close to one, which reflects strong persistence over long horizons. When $\theta < 0$, the first autocovariance in \eqref{acf_arma} is substantially reduced relative to the AR(1) case of $\theta = 0$. A negative $\theta$ combined with a high $\rho$ therefore reconciles a sharp initial decline of the autocorrelation function with strong persistence thereafter.

The two specifications also behave differently under misspecification. If the true model is an AR(1) and we fit the ``rough'' AR model, the estimates are close to the true values, with $\theta$ near zero. If the true model is the ``rough'' AR and we fit an AR(1), the estimated autoregressive coefficient is biased downward. The fitted coefficient converges to the first autocorrelation of the data, which lies below $\rho$ whenever $\theta$ is negative. Section \ref{sec:empirical} documents exactly this pattern. The estimated autoregressive coefficient is systematically higher under the ``rough'' AR model than under the AR(1) model, and $\hat{\theta}$ is negative for every asset. This suggests that the ``rough'' AR specification is the closer of the two to the underlying data-generating process.

This model specification offers several advantages over continuous-time fractional models. First, it is linear and involves only four parameters, $\omega$, $\rho$, $\theta$, and $\sigma$. Being a standard model in time series analysis, it allows the use of a wide range of established estimation techniques. Second, the model is easily extendable, not only through time-varying parameters, structural breaks, and regime switching, but also by incorporating explanatory variables to better capture the dynamics of volatility. This simple structure addresses the two barriers discussed in Section \ref{subsec:limitations}.

An AR(1) structure may nevertheless be too simple to capture the persistence of volatility. We therefore replace the single autoregressive lag with the HAR structure of \citet{corsi:09a}, which approximates the slow decay of the autocorrelations by a sum of autoregressive terms over three time scales:
\begin{align}
y_j = \omega + \rho_1 y_{j-1} + \rho_2 y_{j-1|j-5} + \rho_3 y_{j-1|j-22} + \epsilon_j, \label{har-spec}
\end{align}
where $y_{j-i|j-l} := \frac{1}{l-i+1}\sum_{p=i}^l y_{j-p}$ and $\epsilon_j = \varepsilon_j + \theta \varepsilon_{j-1}$. The key difference between the ``rough'' HAR and the standard HAR model is that in the former, $\epsilon_j$ follows an MA(1) process. When $\theta$ is negative, the model generates roughness and may be referred to as the \emph{``rough'' HAR model}.\footnote{The same caveat applies here. The ``rough'' HAR model imitates the autocovariance behavior of rough processes at the daily frequency without being rough in the sense of Definition \ref{def:rough}.} Equipped with the MA(1) component, the ``rough'' HAR model can capture both long-range dependence and the sharp decline in the first-order ACF, just as the ``rough'' AR model does.

The two components of the model line up with two accounts of how markets operate. The autoregressive part follows \citet{corsi:09a}, who motivates the daily, weekly, and monthly averages by the heterogeneity of trading horizons. Long-term investors, portfolio managers, and day traders react to volatility at different speeds, and the sum of their reactions produces its persistence. The MA(1) term connects to the microstructural reasons for roughness in Section \ref{subsec:ctmodels}, where the splitting of large orders under high endogeneity leaves an anti-persistent signature in volatility at short time scales. Neither argument is a formal derivation, but should be seen as a heuristic motivation, similar to that of the original HAR model. 

\section{Simulation Studies}
\label{sec:simulation}

In this section, we examine the quality of the approximation in a controlled setting. We first illustrate that the rough AR and rough HAR models generate realistic sample paths. We then ask two quantitative questions. Does maximum likelihood recover the mapping between the MA coefficient and the Hurst parameter when the data are generated by a rough continuous-time model? 

\begin{figure}[p]
\begin{center}
\caption{Simulated sample paths of the four models.}
\label{simu_path_com}
\makebox[\textwidth][c]{%
    \includegraphics[width=.9\linewidth]{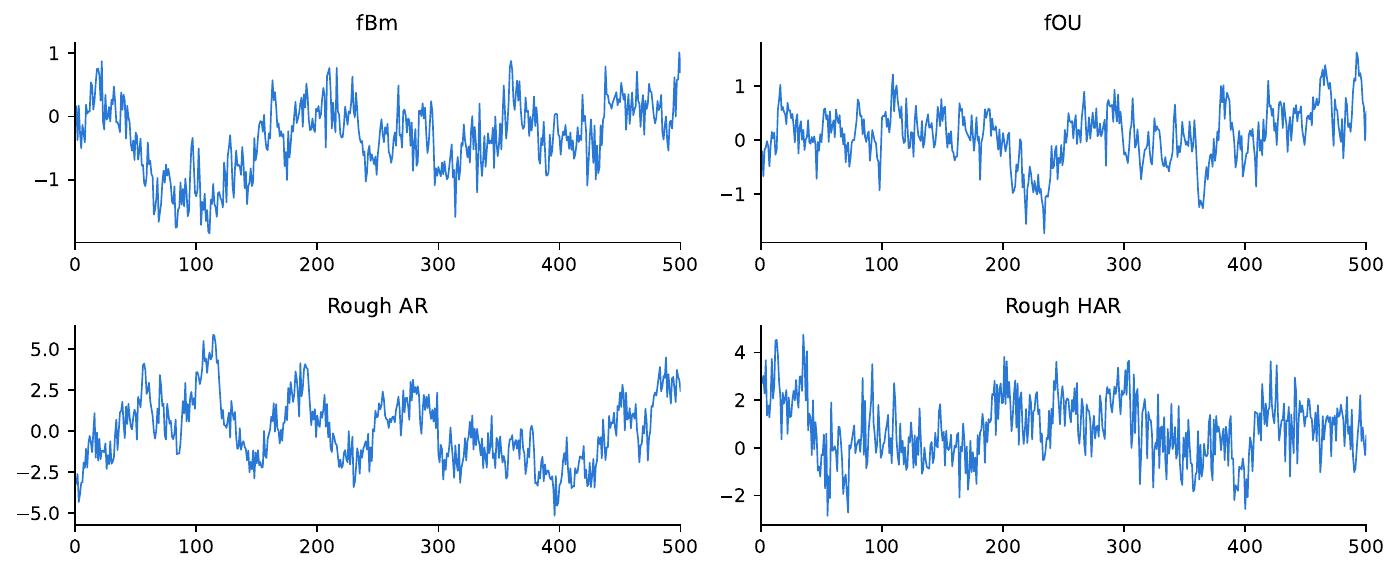}%
}
\begin{scriptsize}
\parbox{\textwidth}{\emph{Note.} We simulate $500$ daily observations from
each model, with $\sigma = 1$ throughout. For fBm and the fOU process,
$H = 0.2$ and $\Delta = 1/250$, with $\kappa = 2$ and $\mu = 0$ for the
latter. For the ``rough'' AR model, $\omega = 0$, $\rho = 0.95$, and
$\theta = -0.4$. For the ``rough'' HAR model, $\omega = 0$,
$(\rho_1, \rho_2, \rho_3) = (0.70, 0.15, 0.10)$, and $\theta = -0.4$. The
autoregressive parameters and the rough AR value of $\theta$ are in line
with the estimates in Table \ref{est_par} and with the rough volatility
literature \citep{gatheral-jaisson-rosenbaum:18a}. We use the same $\theta$
for both rough models, so that the panels differ only in the autoregressive
structure.}
\end{scriptsize}
\end{center}
\end{figure}

\subsection{Sample paths and autocorrelations}
\label{subsec:sim_paths}

\begin{figure}[p]
\begin{center}
\caption{Sample autocorrelation functions of the four models.}
\label{simu_acf_com}
\makebox[\textwidth][c]{%
    \includegraphics[width=.9\linewidth]{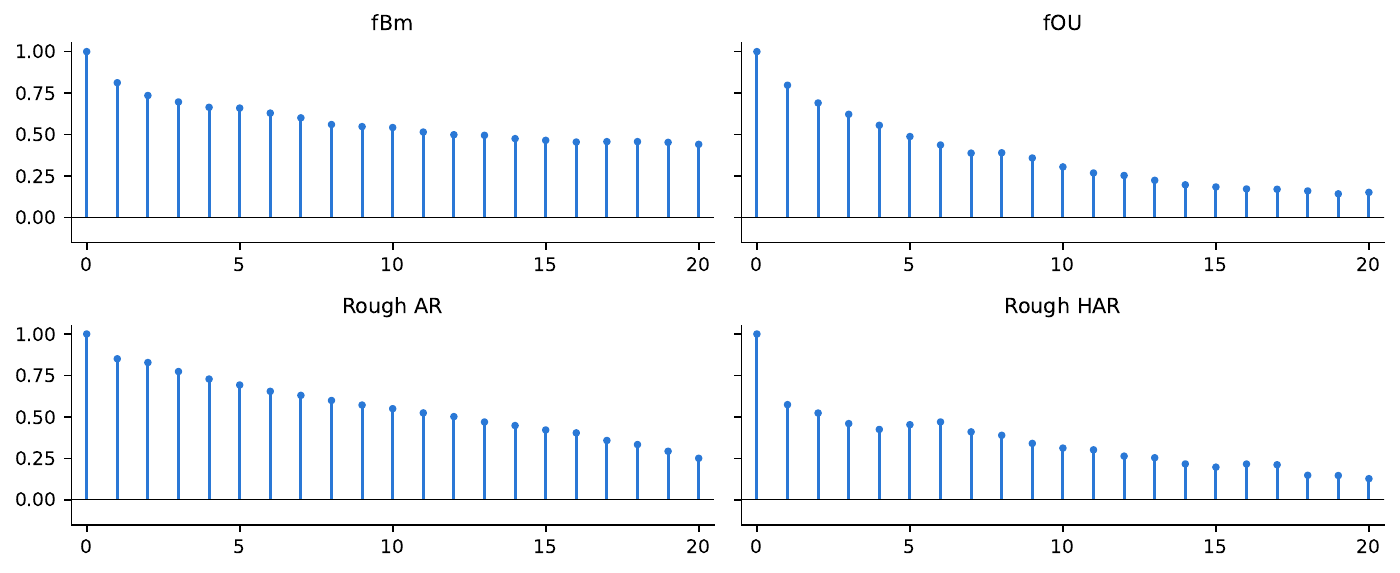}%
}
\begin{scriptsize}
\parbox{\textwidth}{\emph{Note.} The panels show the sample autocorrelation
function of one simulated path from each model. The models and parameter
values are as in Figure \ref{simu_path_com}.}
\end{scriptsize}
\end{center}
\end{figure}

We simulate the four models at parameter values in line with the estimates
of Section \ref{sec:empirical}, as detailed in the note to Figure
\ref{simu_path_com}. Figure \ref{simu_path_com} shows that the rough AR and
rough HAR models replicate the jagged appearance of the fBm and fOU sample
paths. Figure \ref{simu_acf_com} shows that they also capture the slow decay
of the autocorrelation function.

\subsection{Recovering the Hurst parameter}
\label{subsec:sim_recovery}

We next simulate rough continuous-time models and fit the rough AR and rough HAR models to the artificial data. Section \ref{subsec:arma_motivation} suggests that the estimated MA coefficient should track the first Wold coefficient $c_1(H)$. The Monte Carlo experiment examines whether this holds when the autoregressive parameters are estimated jointly.

We consider two data-generating processes. The first is fBm in its unit-root representation, where the increments are fGn. The second is the fOU process sampled at $\Delta = 1/250$ with $\kappa \in \{2, 5\}$. The smaller value is in the middle of our empirical estimates in Table \ref{est_par} and is used in the illustration above. The larger value lies within the range of $\kappa$ used in the simulation study of \citet{bolko-christensen-pakkanen-veliyev:23a}. We simulate the fOU process on a fine grid with ten sub-steps per day, using the autoregressive discretization in \eqref{eq:fou_discrete} with a midpoint approximation of the innovation integral, and we discard a burn-in period of $500$ days so that the recorded sample is stationary. For both processes, the Hurst parameter varies over $H \in \{0.1, 0.2, 0.3, 0.4\}$, which spans the range of empirical estimates in the rough volatility literature \citep{gatheral-jaisson-rosenbaum:18a,bolko-christensen-pakkanen-veliyev:23a}.\footnote{We repeated the experiment on the finer grid $H \in \{0.05, 0.10, \ldots, 0.45\}$. The estimates track $c_1(H)$ along the entire curve.} The sample sizes are $n \in \{500, 4{,}000\}$. The smallest matches the rolling estimation window in Section \ref{sec:empirical}, while the largest is close to the full length of our ETF sample. We run $500$ replications per configuration. The rough AR and rough HAR models are estimated by exact maximum likelihood via the Kalman filter.

Table \ref{tab:sim_theta} summarizes the results for fBm and fOU. Three findings stand out. First, for the rough AR model, $\hat{\theta}$ tracks $c_1(H)$ across the entire grid. At $n = 4{,}000$, the Monte Carlo average differs from $c_1(H)$ by at most $0.013$ for fBm. At $n = 500$, the estimates are mildly attenuated toward zero, so the implied Hurst parameter is slightly too high. Second, for the rough HAR model, $\hat{\theta}$ is biased strongly towards zero at all sample sizes. A similar pattern is also evident in our empirical estimates reported in Table \ref{est_par}. This makes sense, as the $c_1(H)$ term is derived under a different autoregressive dynamics, so the weekly and monthly averages absorb part of the short-run dependence. In consequence, the implied Hurst parameter should be read from the rough AR model. Third, mean reversion plays a relatively minor role when the process is highly persistent. The results for $\kappa = 5$ are close and are deferred to Appendix \ref{sec:app:sim}.

\begin{table}[H]
\centering
\caption{Recovering the Hurst parameter: $\hat\theta$ and implied $H$.}
\label{tab:sim_theta}
\scalebox{0.72}{
\begin{tabular}{cccccccccc}
\toprule
 & & \multicolumn{4}{c}{$n=500$} & \multicolumn{4}{c}{$n=4{,}000$} \\
\cmidrule(lr){3-6} \cmidrule(lr){7-10}
 & & \multicolumn{2}{c}{RAR} & \multicolumn{2}{c}{RHAR} & \multicolumn{2}{c}{RAR} & \multicolumn{2}{c}{RHAR} \\
\cmidrule(lr){3-4} \cmidrule(lr){5-6} \cmidrule(lr){7-8} \cmidrule(lr){9-10}
$H$ & $c_1(H)$ & $\hat\theta$ & $H_{\mathrm{imp}}$ & $\hat\theta$ & $H_{\mathrm{imp}}$ & $\hat\theta$ & $H_{\mathrm{imp}}$ & $\hat\theta$ & $H_{\mathrm{imp}}$ \\
\midrule
\multicolumn{10}{c}{Panel A: fBm} \\
\midrule
0.1 & -0.657 & -0.624 (0.070) & 0.117 & -0.390 (0.258) & 0.247 & -0.670 (0.021) & 0.095 & -0.439 (0.082) & 0.213 \\
0.2 & -0.459 & -0.439 (0.057) & 0.212 & -0.274 (0.122) & 0.314 & -0.467 (0.020) & 0.196 & -0.278 (0.045) & 0.309 \\
0.3 & -0.293 & -0.281 (0.054) & 0.308 & -0.166 (0.103) & 0.383 & -0.298 (0.017) & 0.297 & -0.172 (0.035) & 0.380 \\
0.4 & -0.142 & -0.129 (0.048) & 0.409 & -0.080 (0.075) & 0.440 & -0.142 (0.017) & 0.400 & -0.081 (0.028) & 0.442 \\
\midrule
\multicolumn{10}{c}{Panel B: fOU ($\kappa=2$)} \\
\midrule
0.1 & -0.657 & -0.606 (0.074) & 0.125 & -0.398 (0.255) & 0.243 & -0.628 (0.023) & 0.113 & -0.425 (0.079) & 0.221 \\
0.2 & -0.459 & -0.424 (0.059) & 0.220 & -0.266 (0.132) & 0.319 & -0.440 (0.021) & 0.211 & -0.274 (0.044) & 0.312 \\
0.3 & -0.293 & -0.268 (0.054) & 0.317 & -0.164 (0.101) & 0.384 & -0.280 (0.017) & 0.308 & -0.168 (0.032) & 0.383 \\
0.4 & -0.142 & -0.132 (0.049) & 0.407 & -0.079 (0.083) & 0.440 & -0.137 (0.017) & 0.404 & -0.082 (0.028) & 0.442 \\
\bottomrule
\end{tabular}}
\begin{scriptsize}
\parbox{\textwidth}{\emph{Note.} We simulate 500 replications per cell of the fBm and fOU data-generating processes at the daily frequency ($\Delta=1/250$) and estimate the ``rough'' AR (RAR) and ``rough'' HAR (RHAR) models by exact maximum likelihood via the Kalman filter. The table shows the Monte Carlo average of $\hat\theta$ (standard deviation in parentheses) next to the first Wold coefficient $c_1(H)$ of fGn, and the implied Hurst parameter $H_{\mathrm{imp}}$, computed by inverting the $c_1(\cdot)$ mapping at each replication's $\hat\theta$ and averaging, with the inversion restricted to $H \in (0, 1/2)$. Replications in which the likelihood optimization fails to converge, at most six of $500$ per cell, are excluded.}
\end{scriptsize}
\end{table}

\begin{remark}
\label{rem:pseudo_true}
The rough AR and rough HAR models are misspecified under these data-generating processes. The ML estimator therefore converges to a pseudo-true parameter, which minimizes the Kullback--Leibler divergence between the implied and the true distribution \citep{white:82a}. The pseudo-true value of $\theta$ need not coincide with the Wold coefficient $c_1(H)$, which is derived from the fGn alone. The simulation shows that the two are nevertheless close when the autoregressive part is a single lag. Hence, the truncation argument of Section \ref{subsec:arma_motivation} survives joint estimation.
\end{remark}

\begin{remark}
\label{rem:pileup}
The Monte Carlo standard deviations in Table \ref{tab:sim_theta} are large when $H$ is small, in particular for the rough HAR model at $n = 500$. The MA term and the autoregressive cascade are close substitutes for the short-run dependence, so the model is nearly redundant and $\theta$ is weakly identified. This is no surprise. Section \ref{subsec:arma_motivation} shows that the fOU process is approximately an ARMA(1,1), up to higher-order moving-average terms that are close to zero, so the additional lags of the HAR specification are superfluous under this data-generating process. Many parameter configurations then fit the data almost equally well. In finite samples, the estimate occasionally reaches the invertibility boundary. We treat such fits as spurious, re-estimate from interior starting values, and cap the coefficient just inside the boundary when no interior optimum exists. The forecasts are largely unaffected, since parameter configurations along the flat region of the likelihood imply nearly identical dynamics.
\end{remark}

\begin{table}[H]
\centering
\caption{The ten ETFs.}
\label{table:etfs}
\scalebox{0.7}[0.7]{
\begin{tabular}{ll}
\toprule
Ticker & Fund \\
\midrule
SPY & SPDR S\&P 500 ETF Trust \\
XLB & Materials Select Sector SPDR Fund \\
XLE & Energy Select Sector SPDR Fund \\
XLF & Financial Select Sector SPDR Fund \\
XLI & Industrial Select Sector SPDR Fund \\
XLK & Technology Select Sector SPDR Fund \\
XLP & Consumer Staples Select Sector SPDR Fund \\
XLU & Utilities Select Sector SPDR Fund \\
XLV & Health Care Select Sector SPDR Fund \\
XLY & Consumer Discretionary Select Sector SPDR Fund \\
\bottomrule
\end{tabular}}
\begin{scriptsize}
\parbox{\textwidth}{\emph{Note.} The sample consists of the S\&P 500 market
ETF and the nine Select Sector SPDR funds, which partition the S\&P 500 into
its main industry sectors. For each fund, we compute daily realized variance
and realized quarticity from TAQ data at a five-minute sampling frequency.
The sample runs from January 4, 2010, to December 31, 2024, for a total of roughly
$3{,}770$ trading days.}
\end{scriptsize}
\end{table}

\section{Empirical Studies}
\label{sec:empirical}
We use the daily realized variance of 10 ETFs from January 4, 2010, through December 31, 2024 computed from TAQ data, as listed in Table \ref{table:etfs}. We first provide empirical evidence supporting our model specification and then conduct a horse race for volatility forecasting to demonstrate its usefulness.

\subsection{Evidence for ``rough'' AR and ``rough'' HAR models}
Table \ref{est_par} reports the parameter estimates for all models, applied to the logarithm of RV. We first consider the continuous-time models. The estimated $H$ is below $0.5$ for every asset, and this finding is robust across model specifications. The estimates also imply very high persistence. For fBm, the autoregressive coefficient in the AR(1) representation is exactly $1$. For fOU, the coefficient is $e^{-\kappa\Delta}$, which is around $0.99$ at the estimated values of $\kappa$.

We next turn to the four discrete-time models of Section \ref{sec:motivation}, with the rough AR and rough HAR models abbreviated as RAR and RHAR in the table. Three patterns stand out. First, the MA coefficients are negative for every asset. The classical AR and HAR models thus overlook a salient feature of the data. Second, the autoregressive coefficients are high for both models, and the RAR coefficient is the higher of the two, close to $1$, once the MA term absorbs the roughness. Third, the same shift appears in the HAR models, where accounting for roughness makes the daily coefficient more prominent.

\begin{table}[p]
\centering
\caption{Parameter estimates.}
\label{est_par}
\scalebox{0.64}[0.64]{
\begin{tabular}{lccccccccccccc}
\toprule
 & fBm & \multicolumn{2}{c}{fOU} & AR & \multicolumn{2}{c}{RAR} & \multicolumn{3}{c}{HAR} & \multicolumn{4}{c}{RHAR} \\
\cmidrule(lr){2-2} \cmidrule(lr){3-4} \cmidrule(lr){5-5} \cmidrule(lr){6-7} \cmidrule(lr){8-10} \cmidrule(lr){11-14}
 & $H$ & $H$ & $\kappa$ & $\rho$ & $\rho$ & $\theta$ & $\rho_1$ & $\rho_2$ & $\rho_3$ & $\rho_1$ & $\rho_2$ & $\rho_3$ & $\theta$ \\
\midrule
SPY & 0.1891 & 0.1967 & 2.3479 & 0.8058 & 0.9331 & -0.4111 & 0.5023 & 0.3080 & 0.1134 & 0.6905 & 0.1508 & 0.0969 & -0.1789 \\
 & (0.0076) & (0.0085) & (0.6126) & (0.0091) & (0.0067) & (0.0167) & (0.0187) & (0.0275) & (0.0221) & (0.0455) & (0.0424) & (0.0189) & (0.0425) \\
\addlinespace
XLB & 0.1838 & 0.1887 & 1.4853 & 0.8266 & 0.9470 & -0.4393 & 0.4891 & 0.3196 & 0.1275 & 0.7043 & 0.1409 & 0.1048 & -0.2036 \\
 & (0.0076) & (0.0081) & (0.4739) & (0.0085) & (0.0058) & (0.0159) & (0.0188) & (0.0277) & (0.0217) & (0.0464) & (0.0428) & (0.0184) & (0.0436) \\
\addlinespace
XLE & 0.1813 & 0.1842 & 0.9564 & 0.8546 & 0.9616 & -0.4757 & 0.4694 & 0.3514 & 0.1323 & 0.5572 & 0.2775 & 0.1225 & -0.0817 \\
 & (0.0075) & (0.0079) & (0.3731) & (0.0078) & (0.0046) & (0.0150) & (0.0189) & (0.0281) & (0.0216) & (0.0518) & (0.0486) & (0.0209) & (0.0461) \\
\addlinespace
XLF & 0.1892 & 0.1961 & 2.0288 & 0.8130 & 0.9347 & -0.4076 & 0.5130 & 0.2828 & 0.1334 & 0.6646 & 0.1592 & 0.1164 & -0.1465 \\
 & (0.0076) & (0.0084) & (0.5639) & (0.0083) & (0.0064) & (0.0156) & (0.0187) & (0.0275) & (0.0221) & (0.0462) & (0.0426) & (0.0199) & (0.0434) \\
\addlinespace
XLI & 0.1852 & 0.1910 & 1.8065 & 0.8134 & 0.9402 & -0.4293 & 0.4923 & 0.3173 & 0.1201 & 0.7083 & 0.1375 & 0.0992 & -0.2046 \\
 & (0.0076) & (0.0082) & (0.5282) & (0.0087) & (0.0061) & (0.0159) & (0.0188) & (0.0276) & (0.0220) & (0.0465) & (0.0429) & (0.0185) & (0.0438) \\
\addlinespace
XLK & 0.1762 & 0.1806 & 1.2944 & 0.8034 & 0.9395 & -0.4414 & 0.4783 & 0.3100 & 0.1426 & 0.7037 & 0.1268 & 0.1161 & -0.2121 \\
 & (0.0074) & (0.0079) & (0.4394) & (0.0091) & (0.0063) & (0.0161) & (0.0188) & (0.0280) & (0.0227) & (0.0470) & (0.0426) & (0.0192) & (0.0442) \\
\addlinespace
XLP & 0.1748 & 0.1805 & 1.9327 & 0.7897 & 0.9397 & -0.4601 & 0.4544 & 0.3682 & 0.0942 & 0.7394 & 0.1275 & 0.0731 & -0.2636 \\
 & (0.0074) & (0.0081) & (0.5485) & (0.0081) & (0.0060) & (0.0154) & (0.0191) & (0.0280) & (0.0224) & (0.0497) & (0.0455) & (0.0176) & (0.0469) \\
\addlinespace
XLU & 0.1673 & 0.1720 & 1.6336 & 0.7854 & 0.9431 & -0.4800 & 0.4403 & 0.3704 & 0.1089 & 0.7478 & 0.1161 & 0.0798 & -0.2844 \\
 & (0.0073) & (0.0079) & (0.4994) & (0.0080) & (0.0059) & (0.0158) & (0.0191) & (0.0283) & (0.0228) & (0.0484) & (0.0437) & (0.0178) & (0.0458) \\
\addlinespace
XLV & 0.1844 & 0.1928 & 2.9221 & 0.7934 & 0.9328 & -0.4280 & 0.4870 & 0.3271 & 0.1024 & 0.7225 & 0.1305 & 0.0833 & -0.2230 \\
 & (0.0076) & (0.0085) & (0.6949) & (0.0085) & (0.0067) & (0.0161) & (0.0189) & (0.0276) & (0.0222) & (0.0464) & (0.0427) & (0.0182) & (0.0439) \\
\addlinespace
XLY & 0.1803 & 0.1839 & 1.1264 & 0.8313 & 0.9515 & -0.4549 & 0.4799 & 0.3261 & 0.1364 & 0.6954 & 0.1482 & 0.1111 & -0.2027 \\
 & (0.0076) & (0.0080) & (0.4338) & (0.0087) & (0.0057) & (0.0156) & (0.0189) & (0.0280) & (0.0220) & (0.0488) & (0.0445) & (0.0188) & (0.0457) \\
\bottomrule
\end{tabular}}
\begin{scriptsize}
\parbox{\textwidth}{\emph{Note.} We estimate all models on the logarithm of
daily realized variance for each ETF, with standard errors in parentheses.
The standard errors of the fBm and fOU estimates are computed from the
analytical form of their asymptotic covariance matrices
\citep{fukasawa-takabatake:19a,shi-yu-zhang:24a}, those of the AR and HAR
models from the OLS formulas, that of the RAR model from the outer product
of gradients, and that of the RHAR model from the Hessian of the
likelihood. The sample runs from January 4, 2010, to December 31, 2024. For fBm, we use
the Whittle method of \citet{fukasawa-takabatake:19a}. For fOU, we use the
maximum likelihood method of \citet{wang-xiao-yu-zhang:25a}, with
$\Delta = 1/250$, so that $\kappa$ is expressed per year. The AR and HAR
models are estimated by OLS, and the RAR and RHAR models by maximum
likelihood via the Kalman filter. RAR and RHAR denote the ``rough'' AR model
in \eqref{ar1-spec} and the ``rough'' HAR model in \eqref{har-spec}. An $H$
below $0.5$ indicates roughness, see Definition \ref{def:rough}.}
\end{scriptsize}
\end{table}

\subsection{Forecasting comparison}\label{sec:forecasting_comparison}

We compare the forecasting performance of the models at horizons of one day, one week, and one month. The HAR model of \citet{corsi:09a} serves as the benchmark. We include the log-AR, log-RAR, log-HAR, and log-RHAR models, where the prefix indicates that the model is fitted to the logarithm of RV. The fBm and fOU models represent the continuous-time approach. As a robustness check, we also include three widely used HAR extensions. The HARQ, HARJ, and HARS models augment the HAR model with realized quarticity, jump variation, and realized semivariances, and they  are defined in Appendix \ref{sec:app:model}.

A two-year rolling window of $500$ trading days \citep{gatheral-jaisson-rosenbaum:18a} is employed to fit the models, estimate parameters, and generate $h$-day-ahead forecasts of RV. The window is rolled forward one day at a time, and the parameters are re-estimated at each step. For the models fitted to the logarithm of RV, we convert the forecasts to the level of RV with the log-normal correction, $\widehat{RV}_{t+h} = \exp(\widehat{y}_{t+h} + \widehat{s}_h^{\,2}/2)$, where $\widehat{y}_{t+h}$ is the $h$-step-ahead forecast of log RV and $\widehat{s}_h^{\,2}$ is the model-implied variance of the forecast error. The loss functions used in the analysis include mean squared error (MSE)
and QLIKE \citep{patton:11a, bollerslev-patton-quaedvlieg:16a}, given by 
\begin{eqnarray*}
MSE &=&\left( \widehat{RV}_{t}-RV_{t}\right) ^{2}, \\
QLIKE &=&\frac{RV_{t}}{\widehat{RV}_{t}}-\log \left( \frac{RV_{t}}{\widehat{RV}_{t}} \right)-1.
\end{eqnarray*}
We also compute the $p$-values of the model confidence set (MCS) procedure proposed by \citet{hansen-lunde-nason:11a} to identify the best-performing models, and we report them as stars next to the loss ratios. Appendix \ref{sec:mcs} details the procedure. To mitigate the issue that HAR-type models may occasionally generate implausibly large or small forecasts, we follow \citet{bollerslev-patton-quaedvlieg:16a} and implement an ``insanity filter'' for all forecasts. A forecast outside the range of RV observed in the estimation window is replaced by the mean over that window.

The results are presented in Tables \ref{mse_etf} and \ref{qlike_etf}. Four patterns emerge. First, the logarithmic transformation matters. The log-HAR, log-RAR, and log-RHAR models outperform the HAR benchmark for every ETF at the daily and weekly horizons under both loss functions, as do the fBm and fOU models. Second, the rough models improve on their classical counterparts. The log-RHAR model has a lower QLIKE than the log-HAR model for every ETF at every horizon, and a lower MSE for at least nine of ten ETFs at the daily and weekly horizons. The exception is the monthly horizon under MSE, where the log-HAR model has a small edge and the differences across models are minor. The gain from the MA(1) term is even clearer in the AR family, since the superior performance of the log-AR model relative to the HAR model is not consistent across assets or forecasting horizons, whereas the log-RAR model is consistently strong. Third, the rough discrete-time models are on par with the continuous-time benchmarks. Their average loss ratios are within a few percentage points of those of fBm and fOU at the daily and weekly horizons. At the monthly horizon, the fOU model performs best under QLIKE, while fBm deteriorates under MSE. Fourth, the HAR extensions do not improve on the benchmark in our sample, as the HARQ, HARJ, and HARS models never lower the average loss by more than about half a percent at any horizon.\footnote{These results are consistent with the findings in \cite{shi-yu-zhang:25a}, where such extensions are shown not to improve forecasting performance for either HAR or log-HAR. Accordingly, within the traditional log-HAR family, we report only the results for log-HAR. } The MCS results, reported in full in Appendix \ref{sec:mcs}, sharpen the picture. Under the MSE loss, the test lacks the power to separate the models, and every model survives in the $90\%$ confidence set for every ETF. Under the QLIKE loss, the log-RAR, log-RHAR, fBm, and fOU models are never excluded, whereas the HAR-type models and the log-AR model are excluded for most ETFs at the daily and weekly horizons.


As a robustness check, we repeat the forecasting exercise for a panel of $40$ individual stocks, with data from the VOLARE database system \citep{cipollini-etal:26a}. The findings carry over. The rough models outperform their classical counterparts for nearly every stock, and their accuracy is comparable to that of the continuous-time models. The exception is again the monthly horizon under MSE, where the log-HAR model is ahead for most stocks. The results are reported in Appendix \ref{sec:app:stocks}.

\begin{table}[p]
\centering
\caption{Out-of-sample forecast accuracy, MSE relative to the HAR model.}
\label{mse_etf}
\scalebox{0.7}{
\begin{tabular}{ccccccccccc}
\toprule
  Ticker    & HAR    & HARQ   & HARJ   & HARS   & Log-AR & Log-RAR  & Log-HAR & Log-RHAR & fBm    & fOU    \\
  \midrule
\multicolumn{11}{c}{1-day}  \\
\midrule
SPY & 1.0000\sig{***} & 1.1027\sig{***} & 0.9958\sig{***} & 1.0716\sig{**} & 1.0325\sig{***} & 0.8510\sig{***} & 0.9194\sig{***} & 0.8800\sig{***} & 0.8694\sig{***} & 0.8923\sig{***} \\
XLB & 1.0000\sig{**} & 0.9559\sig{***} & 0.9843\sig{***} & 1.0239\sig{***} & 0.9377\sig{***} & 0.7378\sig{***} & 0.8089\sig{***} & 0.7655\sig{***} & 0.7683\sig{***} & 0.7916\sig{***} \\
XLE & 1.0000\sig{***} & 0.9209\sig{**} & 0.8724\sig{**} & 0.8604\sig{**} & 0.6912\sig{**} & 0.5515\sig{***} & 0.5504\sig{***} & 0.5754\sig{***} & 0.5535\sig{***} & 0.5701\sig{***} \\
XLF & 1.0000\sig{**} & 0.9771\sig{**} & 1.0728\sig{***} & 1.0834\sig{**} & 0.8863\sig{***} & 0.7762\sig{***} & 0.8081\sig{***} & 0.8011\sig{***} & 0.7867\sig{***} & 0.8030\sig{***} \\
XLI & 1.0000\sig{**} & 0.9899\sig{**} & 1.1677\sig{**} & 0.9735\sig{**} & 0.9773\sig{***} & 0.8203\sig{***} & 0.8592\sig{***} & 0.8347\sig{***} & 0.8366\sig{***} & 0.8527\sig{***} \\
XLK & 1.0000\sig{***} & 1.1101\sig{***} & 1.1439\sig{***} & 1.0796\sig{**} & 1.0458\sig{***} & 0.8916\sig{***} & 0.9629\sig{***} & 0.9220\sig{***} & 0.8966\sig{***} & 0.9175\sig{***} \\
XLP & 1.0000\sig{***} & 1.0796\sig{***} & 1.0040\sig{***} & 1.2007\sig{***} & 1.0328\sig{***} & 0.8839\sig{***} & 0.9455\sig{***} & 0.8795\sig{***} & 0.9011\sig{***} & 0.9192\sig{***} \\
XLU & 1.0000\sig{***} & 1.0765\sig{***} & 1.0254\sig{***} & 1.1551\sig{***} & 0.9457\sig{***} & 0.7507\sig{***} & 0.8115\sig{***} & 0.7472\sig{***} & 0.7866\sig{***} & 0.8088\sig{***} \\
XLV & 1.0000\sig{***} & 1.0358\sig{***} & 1.0110\sig{***} & 1.0342\sig{***} & 1.0084\sig{***} & 0.9384\sig{***} & 0.9605\sig{***} & 0.9441\sig{***} & 0.9440\sig{***} & 0.9544\sig{***} \\
XLY & 1.0000\sig{***} & 1.0399\sig{***} & 1.0339\sig{***} & 0.9788\sig{**} & 0.8143\sig{***} & 0.6864\sig{***} & 0.7479\sig{***} & 0.7018\sig{***} & 0.7007\sig{***} & 0.7138\sig{***} \\
\cmidrule(lr){1-11}
\textit{Average} & 1.0000 & 1.0288 & 1.0311 & 1.0461 & 0.9372 & 0.7888 & 0.8374 & 0.8051 & 0.8044 & 0.8223 \\
\midrule
\multicolumn{11}{c}{1-week}  \\
\midrule
SPY & 1.0000\sig{***} & 0.9642\sig{***} & 1.0421\sig{***} & 0.9952\sig{***} & 0.9128\sig{***} & 0.7175\sig{***} & 0.7578\sig{***} & 0.7337\sig{***} & 0.7398\sig{***} & 0.7458\sig{***} \\
XLB & 1.0000\sig{***} & 1.0050\sig{***} & 0.9008\sig{***} & 0.9884\sig{***} & 1.0175\sig{***} & 0.8061\sig{***} & 0.8893\sig{***} & 0.8276\sig{***} & 0.8089\sig{***} & 0.8283\sig{***} \\
XLE & 1.0000\sig{***} & 1.1923\sig{**} & 0.9910\sig{**} & 1.0832\sig{***} & 1.0455\sig{**} & 0.7317\sig{***} & 0.7718\sig{***} & 0.7233\sig{***} & 0.6981\sig{***} & 0.7501\sig{***} \\
XLF & 1.0000\sig{***} & 1.0855\sig{***} & 0.9845\sig{***} & 1.0530\sig{***} & 1.0921\sig{***} & 0.8999\sig{***} & 0.9469\sig{***} & 0.9130\sig{***} & 0.9065\sig{***} & 0.9255\sig{***} \\
XLI & 1.0000\sig{***} & 0.9767\sig{***} & 1.0513\sig{***} & 0.9842\sig{***} & 0.6614\sig{***} & 0.5378\sig{***} & 0.5546\sig{***} & 0.5428\sig{***} & 0.5433\sig{***} & 0.5577\sig{***} \\
XLK & 1.0000\sig{***} & 1.0838\sig{***} & 1.1022\sig{***} & 1.0393\sig{***} & 1.0117\sig{***} & 0.8215\sig{***} & 0.8737\sig{***} & 0.8461\sig{***} & 0.8296\sig{***} & 0.8365\sig{***} \\
XLP & 1.0000\sig{***} & 0.9911\sig{***} & 1.0153\sig{***} & 0.8443\sig{***} & 0.8363\sig{***} & 0.7552\sig{***} & 0.7827\sig{***} & 0.7564\sig{***} & 0.7352\sig{***} & 0.7511\sig{***} \\
XLU & 1.0000\sig{***} & 0.9756\sig{***} & 0.9863\sig{***} & 1.0774\sig{***} & 0.8436\sig{***} & 0.7056\sig{***} & 0.7937\sig{***} & 0.6470\sig{***} & 0.6749\sig{***} & 0.6949\sig{***} \\
XLV & 1.0000\sig{***} & 0.9936\sig{**} & 1.0023\sig{***} & 1.0389\sig{**} & 0.9343\sig{***} & 0.8287\sig{***} & 0.8497\sig{***} & 0.8327\sig{***} & 0.8347\sig{***} & 0.8478\sig{***} \\
XLY & 1.0000\sig{***} & 1.0148\sig{***} & 0.9957\sig{***} & 1.0968\sig{***} & 1.1282\sig{***} & 0.9244\sig{***} & 0.9874\sig{***} & 0.9500\sig{***} & 0.9389\sig{***} & 0.9385\sig{***} \\
\cmidrule(lr){1-11}
\textit{Average} & 1.0000 & 1.0283 & 1.0071 & 1.0201 & 0.9483 & 0.7728 & 0.8208 & 0.7773 & 0.7710 & 0.7876 \\
\midrule
\multicolumn{11}{c}{1-month}  \\
\midrule
SPY & 1.0000\sig{***} & 1.0079\sig{***} & 1.0012\sig{***} & 1.0002\sig{***} & 0.9912\sig{***} & 0.9760\sig{***} & 0.9799\sig{***} & 0.9801\sig{***} & 1.1857\sig{***} & 0.9757\sig{***} \\
XLB & 1.0000\sig{***} & 1.0242\sig{***} & 0.9958\sig{***} & 1.0028\sig{***} & 1.0115\sig{***} & 1.0110\sig{***} & 0.9812\sig{***} & 1.0167\sig{***} & 1.0512\sig{***} & 0.9907\sig{***} \\
XLE & 1.0000\sig{***} & 1.0508\sig{***} & 0.9956\sig{***} & 1.0018\sig{***} & 1.0842\sig{**} & 1.0243\sig{***} & 0.9987\sig{***} & 1.0607\sig{***} & 1.0464\sig{***} & 1.0027\sig{***} \\
XLF & 1.0000\sig{***} & 1.1019\sig{***} & 0.9976\sig{***} & 1.0156\sig{***} & 0.9906\sig{***} & 0.9779\sig{***} & 0.9751\sig{***} & 0.9832\sig{***} & 1.0725\sig{***} & 0.9778\sig{***} \\
XLI & 1.0000\sig{***} & 1.0269\sig{**} & 1.0005\sig{***} & 1.0010\sig{***} & 0.9909\sig{***} & 0.9856\sig{***} & 0.9777\sig{***} & 0.9884\sig{***} & 1.0554\sig{***} & 0.9833\sig{***} \\
XLK & 1.0000\sig{***} & 1.0118\sig{***} & 1.0072\sig{***} & 1.0043\sig{***} & 1.0011\sig{***} & 0.9834\sig{***} & 0.9870\sig{***} & 0.9993\sig{***} & 1.1272\sig{***} & 0.9799\sig{***} \\
XLP & 1.0000\sig{***} & 1.2160\sig{***} & 1.0041\sig{***} & 1.0008\sig{***} & 0.9779\sig{***} & 1.0257\sig{***} & 0.9774\sig{***} & 0.9886\sig{***} & 1.0330\sig{***} & 0.9884\sig{***} \\
XLU & 1.0000\sig{***} & 1.0502\sig{***} & 1.0006\sig{***} & 0.9999\sig{***} & 0.9735\sig{***} & 1.2331\sig{***} & 0.9741\sig{***} & 0.9912\sig{***} & 1.0372\sig{***} & 0.9896\sig{***} \\
XLV & 1.0000\sig{***} & 0.9855\sig{***} & 0.9846\sig{***} & 1.0091\sig{***} & 0.9704\sig{***} & 0.9699\sig{***} & 0.9633\sig{***} & 0.9669\sig{***} & 1.0380\sig{***} & 0.9694\sig{***} \\
XLY & 1.0000\sig{***} & 1.0142\sig{***} & 1.0002\sig{***} & 1.0020\sig{***} & 1.0420\sig{***} & 0.9981\sig{***} & 0.9987\sig{***} & 0.9934\sig{***} & 1.1055\sig{***} & 0.9835\sig{***} \\
\cmidrule(lr){1-11}
\textit{Average} & 1.0000 & 1.0489 & 0.9987 & 1.0037 & 1.0033 & 1.0185 & 0.9813 & 0.9968 & 1.0752 & 0.9841 \\
\bottomrule
\end{tabular}}
\begin{scriptsize}
\parbox{\textwidth}{\emph{Note.}  The
sample runs from January 4, 2010, to December 31, 2024. Each entry is the
ratio of the average MSE of the model in the column to that of the HAR
model, so a value below one indicates an improvement on the benchmark.  The
prefix ``Log'' indicates that the model is fitted to the logarithm of RV,
with forecasts converted to the level of RV by the log-normal correction.
 Superscript stars report the
outcome of the model confidence set (MCS) procedure. Three, two, and one star denote membership
in the $75\%$, $90\%$, and $95\%$ confidence sets, which correspond to MCS
$p$-values above $0.25$, above $0.10$, and above $0.05$. An entry without
stars is excluded from all three sets. The last row of each panel reports
the average ratio across the ten ETFs. The average row carries no stars
because the MCS procedure is applied to each ETF separately.}
\end{scriptsize}
\end{table}

\begin{table}[p]
\centering
\caption{Out-of-sample forecast accuracy, QLIKE relative to the HAR model.}
\label{qlike_etf}
\scalebox{0.7}{
\begin{tabular}{ccccccccccc}
\toprule
  Ticker    & HAR    & HARQ   & HARJ   & HARS   & Log-AR & Log-RAR  & Log-HAR & Log-RHAR & fBm    & fOU    \\
  \midrule
\multicolumn{11}{c}{1-day}  \\
\midrule
SPY & 1.0000\sig{} & 1.0511\sig{} & 1.0226\sig{} & 1.1408\sig{} & 0.9362\sig{} & 0.8695\sig{**} & 0.8769\sig{**} & 0.8676\sig{***} & 0.8622\sig{***} & 0.8594\sig{***} \\
XLB & 1.0000\sig{**} & 1.0638\sig{**} & 1.1280\sig{*} & 1.0721\sig{*} & 0.9604\sig{*} & 0.8725\sig{**} & 0.8893\sig{**} & 0.8695\sig{***} & 0.8590\sig{***} & 0.8634\sig{***} \\
XLE & 1.0000\sig{***} & 1.0483\sig{**} & 0.9434\sig{**} & 0.9262\sig{**} & 0.7893\sig{***} & 0.7085\sig{***} & 0.7035\sig{***} & 0.6981\sig{***} & 0.6901\sig{***} & 0.6943\sig{***} \\
XLF & 1.0000\sig{*} & 0.9698\sig{*} & 1.0741\sig{} & 1.1635\sig{} & 0.8409\sig{*} & 0.7971\sig{***} & 0.8140\sig{**} & 0.8040\sig{***} & 0.7923\sig{***} & 0.7911\sig{***} \\
XLI & 1.0000\sig{} & 1.1973\sig{*} & 0.9953\sig{} & 1.0161\sig{} & 0.9263\sig{*} & 0.8479\sig{***} & 0.8603\sig{**} & 0.8470\sig{***} & 0.8385\sig{***} & 0.8399\sig{***} \\
XLK & 1.0000\sig{*} & 1.0368\sig{*} & 1.0307\sig{*} & 1.1194\sig{} & 0.9202\sig{*} & 0.8537\sig{**} & 0.8602\sig{**} & 0.8512\sig{***} & 0.8444\sig{***} & 0.8442\sig{***} \\
XLP & 1.0000\sig{} & 1.0610\sig{} & 0.9705\sig{} & 1.0869\sig{} & 0.7644\sig{} & 0.6969\sig{***} & 0.7156\sig{**} & 0.6956\sig{***} & 0.6922\sig{***} & 0.6957\sig{***} \\
XLU & 1.0000\sig{**} & 1.0369\sig{**} & 1.0661\sig{*} & 1.1199\sig{*} & 0.9206\sig{**} & 0.8290\sig{***} & 0.8438\sig{**} & 0.8279\sig{***} & 0.8289\sig{***} & 0.8313\sig{***} \\
XLV & 1.0000\sig{} & 0.9671\sig{} & 0.9212\sig{} & 0.9602\sig{} & 0.8312\sig{} & 0.7806\sig{**} & 0.7978\sig{*} & 0.7789\sig{**} & 0.7681\sig{***} & 0.7745\sig{**} \\
XLY & 1.0000\sig{**} & 0.8722\sig{**} & 0.8801\sig{} & 0.9286\sig{} & 0.7515\sig{**} & 0.6862\sig{**} & 0.6928\sig{**} & 0.6857\sig{**} & 0.6729\sig{***} & 0.6764\sig{***} \\
\cmidrule(lr){1-11}
\textit{Average} & 1.0000 & 1.0304 & 1.0032 & 1.0534 & 0.8641 & 0.7942 & 0.8054 & 0.7925 & 0.7849 & 0.7870 \\
\midrule
\multicolumn{11}{c}{1-week}  \\
\midrule
SPY & 1.0000\sig{} & 0.9898\sig{} & 0.9985\sig{} & 0.9944\sig{} & 1.0775\sig{} & 0.8571\sig{***} & 0.8812\sig{***} & 0.8499\sig{***} & 0.8262\sig{***} & 0.8289\sig{***} \\
XLB & 1.0000\sig{} & 0.9804\sig{*} & 0.9605\sig{} & 1.0102\sig{} & 1.1850\sig{} & 0.8572\sig{***} & 0.8756\sig{***} & 0.8448\sig{***} & 0.8088\sig{***} & 0.8270\sig{***} \\
XLE & 1.0000\sig{**} & 1.1811\sig{**} & 1.0526\sig{*} & 1.0319\sig{**} & 1.1535\sig{} & 0.7776\sig{**} & 0.7726\sig{**} & 0.7354\sig{***} & 0.7117\sig{***} & 0.7363\sig{***} \\
XLF & 1.0000\sig{*} & 1.0300\sig{*} & 1.0089\sig{*} & 1.0256\sig{*} & 1.0286\sig{*} & 0.8486\sig{***} & 0.8661\sig{***} & 0.8428\sig{***} & 0.8411\sig{***} & 0.8288\sig{***} \\
XLI & 1.0000\sig{} & 0.9889\sig{} & 0.9772\sig{} & 1.0124\sig{} & 1.1248\sig{} & 0.8502\sig{***} & 0.8686\sig{***} & 0.8425\sig{***} & 0.8176\sig{***} & 0.8228\sig{***} \\
XLK & 1.0000\sig{*} & 0.9459\sig{*} & 1.0079\sig{*} & 0.9483\sig{} & 1.0337\sig{} & 0.8230\sig{***} & 0.8337\sig{***} & 0.8071\sig{***} & 0.7787\sig{***} & 0.8021\sig{***} \\
XLP & 1.0000\sig{} & 1.0852\sig{*} & 0.9880\sig{} & 1.0258\sig{} & 1.0001\sig{} & 0.7521\sig{***} & 0.7857\sig{**} & 0.7575\sig{***} & 0.7289\sig{***} & 0.7430\sig{***} \\
XLU & 1.0000\sig{**} & 1.0389\sig{**} & 0.9953\sig{**} & 1.0661\sig{**} & 1.0078\sig{**} & 0.7387\sig{***} & 0.7740\sig{**} & 0.7301\sig{***} & 0.7340\sig{***} & 0.7485\sig{***} \\
XLV & 1.0000\sig{} & 0.9927\sig{} & 0.9872\sig{} & 0.9890\sig{} & 1.0474\sig{} & 0.8227\sig{***} & 0.8401\sig{***} & 0.8099\sig{***} & 0.7758\sig{***} & 0.8038\sig{***} \\
XLY & 1.0000\sig{} & 0.9912\sig{} & 0.9859\sig{} & 1.0101\sig{} & 1.1374\sig{} & 0.8641\sig{**} & 0.8775\sig{**} & 0.8521\sig{**} & 0.8066\sig{***} & 0.8270\sig{**} \\
\cmidrule(lr){1-11}
\textit{Average} & 1.0000 & 1.0224 & 0.9962 & 1.0114 & 1.0796 & 0.8191 & 0.8375 & 0.8072 & 0.7829 & 0.7968 \\
\midrule
\multicolumn{11}{c}{1-month}  \\
\midrule
SPY & 1.0000\sig{} & 1.0087\sig{} & 0.9852\sig{} & 0.9731\sig{} & 0.9348\sig{***} & 0.8779\sig{***} & 0.9053\sig{***} & 0.8744\sig{***} & 0.8871\sig{***} & 0.8416\sig{***} \\
XLB & 1.0000\sig{**} & 1.0150\sig{*} & 0.9781\sig{***} & 1.0063\sig{*} & 1.0524\sig{***} & 0.9669\sig{***} & 0.9559\sig{***} & 0.9291\sig{***} & 0.9451\sig{***} & 0.9177\sig{***} \\
XLE & 1.0000\sig{**} & 1.0518\sig{**} & 1.0021\sig{**} & 0.9996\sig{**} & 1.2280\sig{} & 1.0089\sig{**} & 0.9129\sig{**} & 0.8949\sig{***} & 0.8625\sig{***} & 0.8925\sig{***} \\
XLF & 1.0000\sig{**} & 1.0320\sig{**} & 0.9876\sig{***} & 1.0021\sig{**} & 0.9794\sig{***} & 0.9300\sig{***} & 0.9706\sig{***} & 0.9592\sig{***} & 1.0132\sig{***} & 0.9211\sig{***} \\
XLI & 1.0000\sig{} & 1.0235\sig{} & 1.0304\sig{} & 0.9946\sig{} & 0.9685\sig{***} & 0.9064\sig{***} & 0.9150\sig{***} & 0.9013\sig{***} & 0.9460\sig{***} & 0.8748\sig{***} \\
XLK & 1.0000\sig{} & 0.9916\sig{} & 0.9841\sig{} & 0.9938\sig{} & 0.9655\sig{*} & 0.9070\sig{***} & 0.9289\sig{***} & 0.8756\sig{***} & 0.8708\sig{***} & 0.8721\sig{***} \\
XLP & 1.0000\sig{***} & 1.0265\sig{} & 1.0173\sig{**} & 1.0001\sig{***} & 0.9352\sig{***} & 0.9341\sig{***} & 0.9548\sig{***} & 0.9463\sig{***} & 1.0174\sig{***} & 0.9205\sig{***} \\
XLU & 1.0000\sig{***} & 1.0083\sig{} & 1.0072\sig{**} & 0.9997\sig{***} & 0.8718\sig{***} & 0.8976\sig{***} & 0.9926\sig{***} & 0.9615\sig{***} & 1.0266\sig{***} & 0.9521\sig{***} \\
XLV & 1.0000\sig{*} & 0.9780\sig{} & 1.0039\sig{**} & 0.9768\sig{} & 0.9147\sig{***} & 0.8690\sig{***} & 0.8820\sig{***} & 0.8544\sig{***} & 0.8669\sig{***} & 0.8449\sig{***} \\
XLY & 1.0000\sig{**} & 1.0151\sig{**} & 1.0149\sig{**} & 1.0014\sig{**} & 1.0541\sig{**} & 0.9517\sig{***} & 0.9565\sig{***} & 0.9132\sig{***} & 0.9124\sig{***} & 0.9013\sig{***} \\
\cmidrule(lr){1-11}
\textit{Average} & 1.0000 & 1.0151 & 1.0011 & 0.9948 & 0.9904 & 0.9250 & 0.9374 & 0.9110 & 0.9348 & 0.8939 \\
\bottomrule
\end{tabular}}
\begin{scriptsize}
\parbox{\textwidth}{\emph{Note.} Each entry is the ratio of the average
QLIKE of the model in the column to that of the HAR model, so a value below
one indicates an improvement on the benchmark. The forecasting design, the
star notation, and the average row are described in the note to Table
\ref{mse_etf}.}
\end{scriptsize}
\end{table}

\subsection{Roughness or measurement error?}
\label{subsec:identification}


RV is an estimate of the latent integrated variance, so log RV measures the latent log variance with error \citep{barndorff-nielsen-shephard:02a,meddahi:03a,andersen-bollerslev-meddahi:04a}. Suppose the latent series follows an AR(1) process and the measurement error is serially independent:
\begin{align}
\label{logrv}
y_j = x_j + w_j, \qquad x_j = \omega + \phi x_{j-1} + v_j,
\end{align}
where $y_j$ is the observed log RV, $x_j$ is the latent log variance, and $(w_j, v_j)$ are mutually independent white-noise sequences. Substituting the first equation into the second yields
\begin{align*}
y_j = \omega + \phi y_{j-1} + u_j, \qquad u_j = v_j + w_j - \phi w_{j-1}.
\end{align*}
The composite error $u_j$ is serially correlated at the first lag only, with autocovariance $-\phi \sigma_w^2 < 0$. The observed series therefore follows an ARMA(1,1) with a negative MA coefficient. This signal-plus-noise model is studied by \citet{hansen-lunde:14a}, who estimate the persistence of the latent process by instrumental variables. 

The measurement errors can generate negative serial correlation. This raises the question of whether noise alone can fully explain the roughness observed in realized variance. We provide two pieces of evidence suggesting that it cannot. The first concerns the forecasting results. If measurement error is the only source of the negative MA coefficient, models designed to exploit measurement error should reap similar forecast gains. The HARQ model of \citet{bollerslev-patton-quaedvlieg:16a} uses realized quarticity to adjust the HAR dynamics for the time-varying magnitude of the measurement error. In Tables \ref{mse_etf} and \ref{qlike_etf}, the HARQ model delivers none of the gains of the rough models. Under the noise reading, this is difficult to explain. Under the roughness reading, it is expected. 

The second piece of evidence uses the asymptotic theory for RV to estimate $\theta$ in the presence of noise. The reduced form above shows that the noise variance and a genuine MA component in the latent process cannot be separated from the dynamics of log RV alone. This can however be resolved by invoking the asymptotic theory for RV. By Theorem 2.1 of \citet{fukasawa-takabatake-westphal:22a}, the errors $w_j$ are asymptotically independent across days and normally distributed with variance $2/m$, where $m$ is the number of intraday returns. At the five-minute frequency over a $6.5$-hour trading day, $m = 78$ and the variance is $1/39$.

\begin{figure}[H]
    \centering
    \includegraphics[width=.7\linewidth]{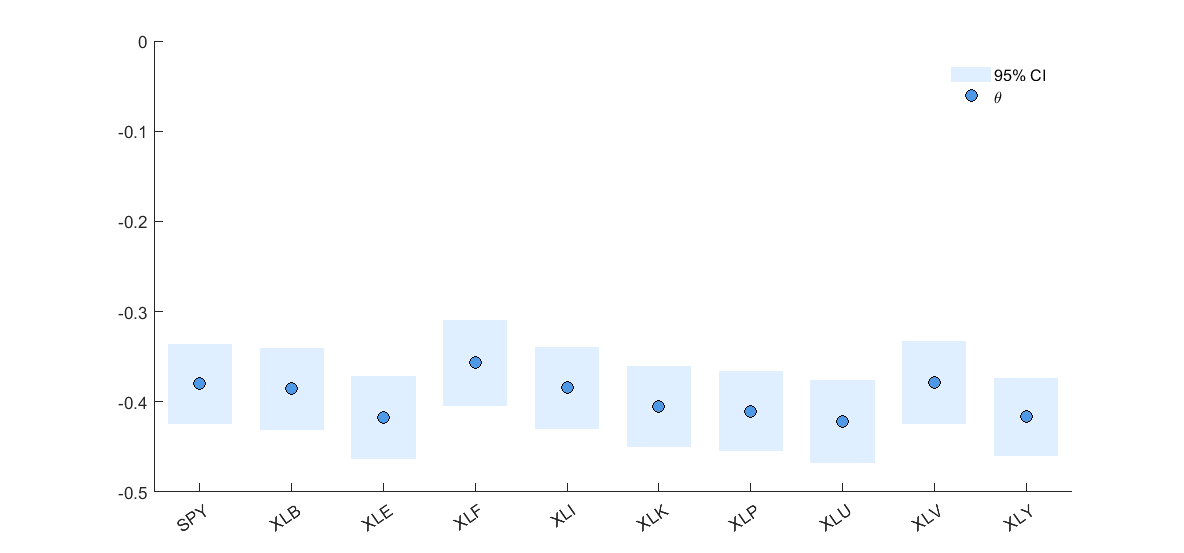}
    \caption{Estimates of $\theta$ and 95\% CI for each asset, accounting for measurement error.}
    \label{theta_noise}
\end{figure}

We therefore fix $\sigma_w^2 = 1/39$ and let the latent innovations follow an MA(1) process,
\begin{align}
\label{logrv-noise}
y_j = x_j + w_j, \qquad x_j = \omega + \phi x_{j-1} + v_j + \theta v_{j-1}.
\end{align}
This is the ``rough'' AR model augmented with measurement noise of known variance. Fixing $\sigma_w^2$ restores identification, and the noise-only explanation becomes the testable restriction $\theta = 0$. We estimate \eqref{logrv-noise} by maximum likelihood via the Kalman filter. Figure \ref{theta_noise} reports the estimates of $\theta$ with $95\%$ confidence intervals. The estimates remain around $-0.4$ for every ETF, slightly attenuated relative to the ``rough'' AR estimates in Table \ref{est_par}, since the noise now absorbs part of the negative correlation. Measurement error of the magnitude implied by the asymptotic theory therefore explains only part of the estimated MA coefficients.



\section{Conclusion}
\label{sec:conclusion}

In this paper, we propose simple ARMA-type models as discrete-time approximations to rough continuous-time models of realized variance.  The approach rests on the observation that an MA(1) model with a negative coefficient provides a good approximation to fGn when the sample path is rough, with the MA(1) coefficient mapping almost linearly into the Hurst parameter. Through a simulation study, we document that fitting the ``rough'' AR model by maximum likelihood to data from a continuous-time fractional process recovers the negative MA coefficient as expected from the Wold expansion, and that the estimates closely track the predicted values. The empirical evidence also supports the approximation. The estimated MA coefficients are negative for every ETF, and the implied Hurst parameters obtained from the ``rough'' AR are in line with those of the continuous-time models. In forecasting, the ``rough'' models outperform their classical counterparts for nearly every asset and horizon, with the largest gains at short horizons, and their accuracy is comparable to that of the rough continuous-time models. The simplicity of the framework leaves ample scope for extensions, such as additional predictors, time-varying parameters, and multivariate systems. We leave these for future research.

\section{Disclosure statement}\label{disclosure-statement}

The authors declare no conflicts of interest.

\section{Data Availability Statement}\label{data-availability-statement}

The data are from VOLARE and are publicly available at https://volare.unime.it/.

\small
\bibliographystyle{rfs}
\bibliography{userref}

 \newpage
\phantomsection\label{supplementary-material}
\bigskip

\begin{center}

{\large\bf SUPPLEMENTARY MATERIAL}

\end{center}
\appendix
\begin{enumerate}
\item Appendix \ref{sec:app:model} collects the definitions of the HAR extensions.
\item Appendix \ref{sec:mcs} summarizes the key procedures of the model confidence set test.
\item Appendix \ref{sec:app:stocks} summarizes the forecasting results for individual stocks.
\item  Appendix \ref{sec:app:sim} reports the Hurst parameter recovery results for the fOU process with alternative $\kappa$.
\item Appendix \ref{app:szego_factor} show how to recovering Wold coefficients from the spectral density via Szegö factorization.
\end{enumerate}

\section{Models}
\label{sec:app:model}
Here we list three more competing models commonly used as benchmarks in the literature.
\subsection*{The HARJ Model}
The HAR-Jump (HARJ) model includes a measure of the jump variation as an additional explanatory variable in the standard HAR model,
\begin{equation*}
RV_t = \beta_0 + \beta_1 RV_{t-1} + \beta_2 RV_{t-1|t-5} + \beta_3 RV_{t-1|t-22}  
+ \beta_j J_{t-1} + u_t, 
\end{equation*}
where \( J_t \equiv \max[RV_t - BPV_t, 0] \), and the BPV measure is defined as,  
\begin{equation*}
BPV_t \equiv \mu_1^{-2} \sum_{i=1}^{M-1} \lvert r_{t,i} \rvert \, \lvert r_{t,i+1} \rvert, 
\end{equation*}
with \( \mu_1 = \sqrt{2/\pi} = \mathbb{E}(\lvert Z \rvert) \), \( Z \) a standard normally distributed random variable, \( r_{t,i} \) the \( i \)th intraday return on day \( t \), and \( M \) the number of intraday observations. 
\subsection*{The HARQ Model}
\begin{equation*}
RV_t = \beta_0 + (\beta_1 + \beta_{1Q} RQ_{t-1}^{1/2}) RV_{t-1} + \beta_2 RV_{t-1|t-5} + \beta_3 RV_{t-1|t-22} + u_t, 
\end{equation*}
where realized quarticity (RQ) is defined by
\begin{equation*}
RQ_t \equiv \frac{M}{3} \sum_{i=1}^M r_{t,i}^4.
\end{equation*}

\subsection*{The HARS Model}
The HAR-Semivariance (HARS) considers the asymmetric impact of positive and negative returns on future volatility:
\begin{equation*}
RV_{t+h} = \beta_0 + \beta_1^+ RS_t^+ + \beta_1^- RS_t^- + \beta_2 RV_{t|t-4} + \beta_3 RV_{t|t-21} + \varepsilon_{t+h},
\end{equation*}
where
\begin{equation*}
RS_t^+ = \sum_{i=1}^M r_{t,i}^2 \mathbb{I}\{r_{t,i} > 0\}, \quad RS_t^- = \sum_{i=1}^M r_{t,i}^2 \mathbb{I}\{r_{t,i} < 0\},
\end{equation*}
and \( \mathbb{I}\{\cdot\} \) is the indicator function.


\section{Model confidence set test}
\label{sec:mcs}
The model confidence set (MCS) provides a confidence set containing the best models with a probability
greater than or equal to a specified level (e.g., 90\%). It also assigns a
p-value to each individual model, allowing for a comprehensive evaluation of
their statistical significance.

Suppose we have a set of competing models indexed by $\mathbb{M}%
_{0}=\left\{ {i=1,\ldots ,M}\right\} $. The loss function ($L_{i,t}$) can be squared forecast error  for model $i$ at time $t$. We calculate the relative
performance as $d_{{ij,t}}=L_{i,t}-L_{j,t}$ for all $i,j\in \mathbb{M}_{0}$%
. The MCS procedure involves an iterative process to identify the model confidence set. For iteration $s$, the null and alternative hypotheses are as follows: 
\begin{equation*}
H_{0,\mathbb{M}_{s}}:\E(d_{ij,t})=0\,\text{for\thinspace\ all}\,i,j\in 
\mathbb{M}_{s}\subset \mathbb{M}_{0},
\end{equation*}%
and the alternative 
\begin{equation*}
H_{A,\mathbb{M}_{s}}:\E(d_{ij,t})\neq 0\,\text{for\thinspace\ some}\,i,j\in 
\mathbb{M}_{s}.
\end{equation*}

We perform a model equivalence test using the $T_{\max,M_s}$ statistic with a
bootstrapped implementation (block length of 20 and $5,000$ replications). The $T_{\max,\mathbb{M}_s}$ statistic is defined as the maximum of the studentized average loss differential of each model relative to the other models in $\mathbb{M}_s$.  If $H_{0,%
\mathbb{M}_s}$ is not rejected, the best model confidence set is $\mathbb{M%
}_s$. Otherwise, we apply an elimination rule to remove models from $%
\mathbb{M}_s$ according to the guidelines specified in \citet{hansen-lunde-nason:11a}
and repeat the test.

Let $P_{H_{0,\mathbb{M}_{s}}}$ denote the p-value associated with the null
hypothesis $H_{0,\mathbb{M}_{s}}$, and let $e_{\mathbb{M}_{s}}$ represent
the model eliminated from the set $\mathbb{M}_{s}$ when $H_{0,\mathbb{M}%
_{s}}$ is rejected. The MCS p-value for model $e_{\mathbb{M}_{s}}$ is
defined as

\begin{equation*}
\hat{p}_{e_{\mathbb{M}_{s}}} = \max_{k \leq s} P_{H_{0,\mathbb{M}_{k}}},
\end{equation*}
where $\mathbb{M}_{1} \supset \mathbb{M}_{2} \supset \ldots \supset 
\mathbb{M}_{s}$.

Tables \ref{mse_mcs_etf} and \ref{mcs_qlike_etf} report the MCS $p$-values behind the stars in Tables \ref{mse_etf} and \ref{qlike_etf}.

\begin{table}[H]
\centering
\caption{MCS $p$-values under the MSE loss.}
\label{mse_mcs_etf}
\scalebox{0.72}{
\begin{tabular}{ccccccccccc}
\toprule
  Ticker    & HAR    & HARQ   & HARJ   & HARS   & Log-AR & Log-RAR  & Log-HAR & Log-RHAR & fBm    & fOU    \\
\midrule
  \multicolumn{11}{c}{1-day}  \\
\midrule
SPY & 0.3878 & 0.3878 & 0.5752 & 0.1728 & 0.3980 & 1.0000   & 0.5752  & 0.5780   & 0.6196 & 0.6196 \\
XLB & 0.1882 & 0.3842 & 0.3616 & 0.2816 & 0.3616 & 1.0000   & 0.3842  & 0.6332   & 0.6332 & 0.6332 \\
XLE & 0.2708 & 0.2354 & 0.2354 & 0.2354 & 0.2354 & 0.8974   & 1.0000  & 0.4488   & 0.8974 & 0.5640 \\
XLF & 0.2172 & 0.2094 & 0.3822 & 0.2062 & 0.3822 & 1.0000   & 0.3822  & 0.3822   & 0.5662 & 0.5662 \\
XLI & 0.1622 & 0.2020 & 0.2020 & 0.2020 & 0.3810 & 1.0000   & 0.3810  & 0.6128   & 0.6128 & 0.6128 \\
XLK & 0.3280 & 0.3280 & 0.5226 & 0.1824 & 0.3038 & 1.0000   & 0.5226  & 0.5226   & 0.7046 & 0.5946 \\
XLP & 0.6604 & 0.6158 & 0.6604 & 0.2996 & 0.6384 & 0.6604   & 0.6604  & 1.0000   & 0.6604 & 0.6604 \\
XLU & 0.4054 & 0.4054 & 0.4054 & 0.2670 & 0.4054 & 0.6988   & 0.6196  & 1.0000   & 0.6196 & 0.6196 \\
XLV & 0.3066 & 0.3066 & 0.3066 & 0.3066 & 0.3066 & 1.0000   & 0.3066  & 0.6390   & 0.6390 & 0.6390 \\
XLY & 0.4672 & 0.4672 & 0.3776 & 0.2178 & 0.4672 & 1.0000   & 0.4672  & 0.7180   & 0.7180 & 0.7180\\
\midrule
\multicolumn{11}{c}{1-week}  \\
\midrule
SPY & 0.6410 & 0.3280 & 0.5758 & 0.6636 & 0.6636 & 1.0000   & 0.7178  & 0.9020   & 0.9020 & 0.8502 \\
XLB & 0.7174 & 0.5532 & 0.7174 & 0.7174 & 0.7174 & 1.0000   & 0.7174  & 0.7820   & 0.7820 & 0.7820 \\
XLE & 0.4822 & 0.1804 & 0.2206 & 0.2846 & 0.2206 & 0.5498   & 0.4822  & 0.6482   & 1.0000 & 0.6482 \\
XLF & 0.7184 & 0.3020 & 0.7736 & 0.6744 & 0.6920 & 1.0000   & 0.7736  & 0.8366   & 0.8366 & 0.8366 \\
XLI & 0.4652 & 0.4916 & 0.4916 & 0.4114 & 0.4916 & 1.0000   & 0.4916  & 0.8840   & 0.8840 & 0.7160 \\
XLK & 0.6510 & 0.3026 & 0.5514 & 0.6510 & 0.6510 & 1.0000   & 0.6510  & 0.6510   & 0.8442 & 0.8442 \\
XLP & 0.6934 & 0.3752 & 0.6934 & 0.7198 & 0.7198 & 0.7198   & 0.7198  & 0.7198   & 1.0000 & 0.7198 \\
XLU & 0.6650 & 0.3350 & 0.6600 & 0.6054 & 0.6786 & 0.7534   & 0.6786  & 1.0000   & 0.7534 & 0.7534 \\
XLV & 0.2846 & 0.1634 & 0.2846 & 0.2198 & 0.4376 & 1.0000   & 0.5538  & 0.9268   & 0.9268 & 0.7070 \\
XLY & 0.7756 & 0.5438 & 0.7756 & 0.4888 & 0.5438 & 1.0000   & 0.7756  & 0.7756   & 0.9532 & 0.9532\\
\midrule
\multicolumn{11}{c}{1-month}  \\
\midrule
SPY & 0.3110 & 0.3110 & 0.3110 & 0.3110 & 0.3110 & 0.9764   & 0.8728  & 0.8430   & 0.3110 & 1.0000 \\
XLB & 0.7352 & 0.3528 & 0.7352 & 0.7352 & 0.5982 & 0.6482   & 1.0000  & 0.7352   & 0.5982 & 0.7352 \\
XLE & 0.9768 & 0.5490 & 1.0000 & 0.9538 & 0.2118 & 0.6032   & 0.9768  & 0.6032   & 0.6032 & 0.9768 \\
XLF & 0.5316 & 0.5316 & 0.5316 & 0.5316 & 0.5316 & 0.8536   & 1.0000  & 0.6058   & 0.5210 & 0.8536 \\
XLI & 0.5304 & 0.2338 & 0.5304 & 0.5304 & 0.6922 & 0.6922   & 1.0000  & 0.6922   & 0.5304 & 0.6922 \\
XLK & 0.6860 & 0.4200 & 0.4200 & 0.4962 & 0.6860 & 0.8408   & 0.8408  & 0.6860   & 0.4200 & 1.0000 \\
XLP & 0.7400 & 0.5054 & 0.7400 & 0.7400 & 0.9236 & 0.7400   & 1.0000  & 0.7400   & 0.6382 & 0.7400 \\
XLU & 0.5930 & 0.4430 & 0.5930 & 0.5930 & 1.0000 & 0.5618   & 0.9336  & 0.5930   & 0.5930 & 0.5930 \\
XLV & 0.5916 & 0.5916 & 0.5916 & 0.5916 & 0.7298 & 0.7298   & 1.0000  & 0.7298   & 0.5916 & 0.7298 \\
XLY & 0.9358 & 0.4562 & 0.9260 & 0.8356 & 0.2984 & 0.9358   & 0.9358  & 0.9358   & 0.4562 & 1.0000\\
\bottomrule
\end{tabular}}
\begin{scriptsize}
\parbox{\textwidth}{\emph{Note.} We compute the model confidence set of
\citet{hansen-lunde-nason:11a} for each ETF and horizon, based on the MSE
loss of the forecasts in Table \ref{mse_etf}. Each entry is the MCS
$p$-value of the model in the column. A $p$-value below $0.10$ indicates
that the model is excluded from the $90\%$ confidence set.}
\end{scriptsize}
\end{table}

\begin{table}[H]
\centering
\caption{MCS $p$-values under the QLIKE loss.}
\label{mcs_qlike_etf}
\scalebox{0.72}{
\begin{tabular}{ccccccccccc}
\toprule
  Ticker    & HAR    & HARQ   & HARJ   & HARS   & Log-AR & Log-RAR  & Log-HAR & Log-RHAR & fBm    & fOU    \\
\midrule
  \multicolumn{11}{c}{1-day}  \\
\midrule
SPY & 0.0120 & 0.0284 & 0.0120 & 0.0120 & 0.0284 & 0.2180   & 0.2180  & 0.3490   & 0.4536 & 1.0000 \\
XLB & 0.1148 & 0.1148 & 0.0680 & 0.0888 & 0.0888 & 0.2078   & 0.1148  & 0.2600   & 1.0000 & 0.2600 \\
XLE & 0.2614 & 0.1028 & 0.1324 & 0.1096 & 0.2614 & 0.2614   & 0.2614  & 0.3582   & 1.0000 & 0.3582 \\
XLF & 0.0756 & 0.0944 & 0.0438 & 0.0438 & 0.0944 & 0.3634   & 0.1824  & 0.3634   & 0.7328 & 1.0000 \\
XLI & 0.0160 & 0.0670 & 0.0108 & 0.0108 & 0.0670 & 0.3062   & 0.1006  & 0.3320   & 1.0000 & 0.7276 \\
XLK & 0.0596 & 0.0540 & 0.0596 & 0.0002 & 0.0596 & 0.1988   & 0.1988  & 0.3336   & 0.9658 & 1.0000 \\
XLP & 0.0114 & 0.0380 & 0.0086 & 0.0084 & 0.0380 & 0.7928   & 0.1212  & 0.8462   & 1.0000 & 0.8462 \\
XLU & 0.1322 & 0.1322 & 0.0930 & 0.0504 & 0.1322 & 0.9894   & 0.1322  & 1.0000   & 0.9894 & 0.8492 \\
XLV & 0.0016 & 0.0206 & 0.0016 & 0.0000 & 0.0206 & 0.1152   & 0.0578  & 0.1732   & 1.0000 & 0.2384 \\
XLY & 0.1096 & 0.1096 & 0.0222 & 0.0006 & 0.1096 & 0.1096   & 0.1096  & 0.1096   & 1.0000 & 0.3012\\
\midrule
\multicolumn{11}{c}{1-week}  \\
\midrule
SPY & 0.0104 & 0.0124 & 0.0124 & 0.0026 & 0.0026 & 0.3078   & 0.3078  & 0.4518   & 1.0000 & 0.8428 \\
XLB & 0.0418 & 0.0936 & 0.0232 & 0.0396 & 0.0132 & 0.2672   & 0.2672  & 0.2672   & 1.0000 & 0.2672 \\
XLE & 0.1298 & 0.1338 & 0.0994 & 0.1232 & 0.0008 & 0.1338   & 0.1338  & 0.4000   & 1.0000 & 0.4000 \\
XLF & 0.0518 & 0.0518 & 0.0518 & 0.0518 & 0.0518 & 0.6908   & 0.3482  & 0.7004   & 0.7004 & 1.0000 \\
XLI & 0.0172 & 0.0208 & 0.0208 & 0.0208 & 0.0208 & 0.3356   & 0.3356  & 0.4266   & 1.0000 & 0.7232 \\
XLK & 0.0730 & 0.0730 & 0.0730 & 0.0176 & 0.0176 & 0.2522   & 0.2522  & 0.5402   & 1.0000 & 0.5402 \\
XLP & 0.0130 & 0.0752 & 0.0088 & 0.0226 & 0.0242 & 0.5986   & 0.1378  & 0.5986   & 1.0000 & 0.5986 \\
XLU & 0.1252 & 0.1252 & 0.1252 & 0.1252 & 0.1252 & 0.8428   & 0.2304  & 1.0000   & 0.8428 & 0.7284 \\
XLV & 0.0044 & 0.0044 & 0.0044 & 0.0016 & 0.0044 & 0.2720   & 0.2720  & 0.2720   & 1.0000 & 0.2720 \\
XLY & 0.0074 & 0.0062 & 0.0010 & 0.0074 & 0.0004 & 0.2176   & 0.2176  & 0.2176   & 1.0000 & 0.2176\\
\midrule
\multicolumn{11}{c}{1-month}  \\
\midrule
SPY & 0.0352 & 0.0352 & 0.0204 & 0.0352 & 0.4790 & 0.7840   & 0.4790  & 0.7840   & 0.7840 & 1.0000 \\
XLB & 0.1160 & 0.0694 & 0.4042 & 0.0910 & 0.2802 & 0.8090   & 0.8090  & 0.8090   & 0.8090 & 1.0000 \\
XLE & 0.1918 & 0.1918 & 0.1918 & 0.1918 & 0.0150 & 0.1918   & 0.2228  & 0.3186   & 1.0000 & 0.3574 \\
XLF & 0.1328 & 0.2400 & 0.3962 & 0.1234 & 0.8096 & 0.8096   & 0.8096  & 0.8096   & 0.6978 & 1.0000 \\
XLI & 0.0272 & 0.0272 & 0.0272 & 0.0272 & 0.5472 & 0.7022   & 0.6482  & 0.7022   & 0.6482 & 1.0000 \\
XLK & 0.0446 & 0.0398 & 0.0446 & 0.0446 & 0.0876 & 0.3186   & 0.3110  & 0.9562   & 1.0000 & 0.9774 \\
XLP & 0.3944 & 0.0054 & 0.1232 & 0.3904 & 0.9188 & 0.9188   & 0.8370  & 0.9188   & 0.7068 & 1.0000 \\
XLU & 0.3872 & 0.0370 & 0.2408 & 0.3628 & 1.0000 & 0.7650   & 0.3872  & 0.6112   & 0.5452 & 0.7190 \\
XLV & 0.0902 & 0.0372 & 0.1112 & 0.0186 & 0.4518 & 0.8456   & 0.4518  & 0.8456   & 0.8456 & 1.0000 \\
XLY & 0.1866 & 0.1866 & 0.1424 & 0.1866 & 0.1866 & 0.3848   & 0.3068  & 0.7806   & 0.7910 & 1.0000\\
\bottomrule
\end{tabular}}
\begin{scriptsize}
\parbox{\textwidth}{\emph{Note.} We compute the model confidence set of
\citet{hansen-lunde-nason:11a} for each ETF and horizon, based on the QLIKE
loss of the forecasts in Table \ref{qlike_etf}. Each entry is the MCS
$p$-value of the model in the column. A $p$-value below $0.10$ indicates
that the model is excluded from the $90\%$ confidence set.}
\end{scriptsize}
\end{table}

\section{Individual stocks}
\label{sec:app:stocks}
This appendix reports the forecasting results for the panel of $40$ individual stocks. Table \ref{tab:stock_symbols} describes the panel.
\begin{table}[H]
\centering
\caption{The panel of individual stocks.}
\label{tab:stock_symbols}
\scalebox{0.56}{
\begin{tabular}{lllp{2.5cm}}
\toprule
Symbol & Name & Sector & First date \\
\midrule
AAPL  & Apple Inc. & Information Technology & 2015-01-02 \\
ADBE   & Adobe Inc. & Information Technology & 2015-01-02 \\
AMD    & Advanced Micro Devices & Information Technology & 2015-01-02 \\
AMGN  & Amgen Inc. & Health Care & 2015-01-02 \\
AMZN  & Amazon & Consumer Discretionary & 2015-01-02 \\
AXP  & American Express & Finance & 2015-01-02 \\
BA   & Boeing & Industrials & 2015-01-02 \\
CAT   & Caterpillar Inc. & Industrials & 2015-01-02 \\
CRM   & Salesforce Inc. & Information Technology & 2015-01-02 \\
CSCO  & Cisco & Information Technology & 2015-01-02 \\
CVX   & Chevron Corporation & Energy & 2015-01-02 \\
DIS   & Walt Disney Company (The) & Communication Services & 2015-01-02 \\
GE     & GE Aerospace & Industrials & 2015-01-02 \\
GOOGL  & Alphabet Inc. (Class A) & Communication Services & 2015-01-02 \\
GS    & Goldman Sachs & Finance & 2015-01-02 \\
HD    & Home Depot & Consumer Discretionary & 2015-01-02 \\
HON  & Honeywell International Inc. & Industrials & 2015-01-02 \\
IBM   & IBM & Information Technology & 2015-01-02 \\
JNJ   & Johnson \& Johnson & Health Care & 2015-01-02 \\
JPM   & JPMorgan Chase & Finance & 2015-01-02 \\
KO    & Coca-Cola Company (The) & Consumer Staples & 2015-01-02 \\
MCD   & McDonald's & Consumer Discretionary & 2015-01-02 \\
META   & Meta Platforms & Communication Services & 2015-01-02 \\
MMM   & 3M & Industrials & 2015-01-02 \\
MRK   & Merck \& Company Inc. & Health Care & 2015-01-02 \\
MSFT  & Microsoft & Information Technology & 2015-01-02 \\
NFLX   & Netflix, Inc. & Communication Services & 2015-01-02 \\
NKE   & Nike, Inc. & Consumer Discretionary & 2015-01-02 \\
NVDA  & Nvidia & Information Technology & 2015-01-02 \\
ORCL   & Oracle Corporation & Information Technology & 2015-01-02 \\
PG   & Procter \& Gamble & Consumer Staples & 2015-01-02 \\
PM     & Philip Morris International & Consumer Staples & 2015-01-02 \\
SHW   & Sherwin-Williams Company & Consumer Discretionary & 2015-01-02 \\
TRV   & The Travelers Companies Inc. & Finance & 2015-01-02 \\
TSLA   & Tesla, Inc. & Consumer Discretionary & 2015-01-02 \\
UNH   & Unitedhealth Group Inc. & Health Care & 2015-01-02 \\
V     & Visa Inc. & Finance & 2015-01-02 \\
VZ    & Verizon Communications Inc. & Public Utilities & 2015-01-02 \\
WMT   & Walmart & Consumer Staples & 2015-01-02 \\
XOM    & ExxonMobil & Energy & 2015-01-02 \\
\bottomrule
\end{tabular}}
\begin{scriptsize}
\parbox{\textwidth}{\emph{Note.} The table lists the $40$ stocks in the
panel, with company name, sector, and the first date of the sample. The
data are from the VOLARE database system \citep{cipollini-etal:26a},
available at \url{http://volare.unime.it}.}
\end{scriptsize}
\end{table}

\subsection*{One-day-ahead forecasts}
\begin{table}[H]
\centering
\caption{Out-of-sample forecast accuracy for individual stocks, MSE relative to the HAR model, one-day horizon.}
\label{mse_stock_1d}
\scalebox{0.66}{
\begin{tabular}{lcccccccccc}
\toprule
  Ticker      & HAR    & HARQ   & HARJ   & HARS   & Log-AR & Log-RAR  & Log-HAR & Log-RHAR & fBm    & fOU    \\
\midrule
AAPL  & 1.0000 & 0.9511 & 0.9838 & 1.0117 & 0.8903 & 0.8143   & 0.8542 & 0.8245  & 0.8058 & 0.8166 \\
ADBE  & 1.0000 & 1.1572 & 1.0556 & 1.0077 & 0.8989 & 0.8135   & 0.8765 & 0.8341  & 0.8204 & 0.8300 \\
AMD   & 1.0000 & 0.9285 & 1.0060 & 0.9997 & 0.9550 & 0.9040   & 0.9154 & 0.9104  & 0.9005 & 0.8995 \\
AMGN  & 1.0000 & 1.0740 & 1.0234 & 1.0992 & 0.8117 & 0.7108   & 0.7559 & 0.7419  & 0.6857 & 0.7174 \\
AMZN  & 1.0000 & 0.9466 & 0.9427 & 1.0519 & 1.0025 & 0.9304   & 0.9301 & 0.9303  & 0.9282 & 0.9289 \\
AXP   & 1.0000 & 1.0141 & 0.9894 & 0.7553 & 0.8507 & 0.6712   & 0.7049 & 0.7025  & 0.6687 & 0.6907 \\
BA    & 1.0000 & 0.8798 & 1.0391 & 1.1976 & 1.0481 & 0.8344   & 0.8852 & 0.8582  & 0.7994 & 0.8452 \\
CAT   & 1.0000 & 1.2719 & 0.9921 & 1.0894 & 1.0805 & 0.8948   & 0.9742 & 0.9339  & 0.9198 & 0.9377 \\
CRM   & 1.0000 & 0.9380 & 1.0849 & 0.9956 & 0.9828 & 0.9080   & 0.9431 & 0.9178  & 0.9057 & 0.9154 \\
CSCO  & 1.0000 & 0.9604 & 1.4584 & 1.0521 & 1.0127 & 0.8202   & 0.8935 & 0.8571  & 0.7962 & 0.8476 \\
CVX   & 1.0000 & 0.9232 & 1.0206 & 0.9295 & 0.6552 & 0.4883   & 0.5212 & 0.5231  & 0.5012 & 0.5224 \\
DIS   & 1.0000 & 1.2103 & 0.9041 & 1.0295 & 1.1276 & 0.8669   & 0.8882 & 0.8856  & 0.8103 & 0.8861 \\
GE    & 1.0000 & 0.9550 & 1.1354 & 1.0090 & 1.0408 & 0.9044   & 0.9662 & 0.9521  & 0.9012 & 0.9124 \\
GOOGL & 1.0000 & 0.9841 & 0.9921 & 1.0194 & 1.0149 & 0.9544   & 0.9655 & 0.9635  & 0.9526 & 0.9554 \\
GS    & 1.0000 & 1.0799 & 1.0143 & 0.9601 & 0.5919 & 0.5118   & 0.5517 & 0.5319  & 0.5102 & 0.5265 \\
HD    & 1.0000 & 0.9836 & 1.1168 & 1.0988 & 0.8000 & 0.6545   & 0.6816 & 0.7172  & 0.6584 & 0.6845 \\
HON   & 1.0000 & 0.8123 & 0.9436 & 1.0390 & 0.9152 & 0.7463   & 0.8370 & 0.8014  & 0.7327 & 0.7692 \\
IBM   & 1.0000 & 1.0357 & 1.1392 & 1.0423 & 0.7879 & 0.6394   & 0.7035 & 0.6855  & 0.6379 & 0.6629 \\
JNJ   & 1.0000 & 1.1690 & 0.9946 & 1.0935 & 1.3309 & 1.0010   & 1.0717 & 1.0047  & 1.0087 & 1.0672 \\
JPM   & 1.0000 & 1.1298 & 0.9780 & 0.9164 & 0.9644 & 0.7842   & 0.8100 & 0.8019  & 0.7862 & 0.8163 \\
KO    & 1.0000 & 1.0993 & 1.4322 & 1.5219 & 1.3032 & 0.9688   & 1.0497 & 0.9779  & 1.0088 & 1.0515 \\
MCD   & 1.0000 & 1.0016 & 0.9834 & 0.7237 & 0.4515 & 0.3491   & 0.3813 & 0.3769  & 0.3401 & 0.3622 \\
META  & 1.0000 & 0.9905 & 0.9758 & 1.0533 & 0.8453 & 0.7676   & 0.7997 & 0.7843  & 0.7536 & 0.7664 \\
MMM   & 1.0000 & 1.0136 & 1.1117 & 1.1343 & 1.1007 & 0.9285   & 0.9800 & 0.9467  & 0.9134 & 0.9491 \\
MRK   & 1.0000 & 1.2214 & 1.0260 & 1.0799 & 1.1641 & 0.9441   & 1.0060 & 0.9726  & 0.9379 & 0.9741 \\
MSFT  & 1.0000 & 1.0718 & 1.0263 & 1.0622 & 1.0101 & 0.9369   & 0.9730 & 0.9589  & 0.9260 & 0.9365 \\
NFLX  & 1.0000 & 0.9622 & 0.9532 & 1.0146 & 0.9932 & 0.9332   & 0.9370 & 0.9353  & 0.9314 & 0.9321 \\
NKE   & 1.0000 & 0.9721 & 1.0377 & 0.8808 & 0.8442 & 0.6955   & 0.7412 & 0.7208  & 0.7009 & 0.7197 \\
NVDA  & 1.0000 & 0.9282 & 1.0019 & 0.9987 & 0.9645 & 0.8912   & 0.9408 & 0.9028  & 0.8990 & 0.8982 \\
ORCL  & 1.0000 & 0.9969 & 1.1687 & 0.9805 & 0.9115 & 0.8137   & 0.8446 & 0.8277  & 0.7999 & 0.8237 \\
PG    & 1.0000 & 1.0876 & 1.1733 & 0.9178 & 0.9738 & 0.6900   & 0.7690 & 0.7064  & 0.7179 & 0.7531 \\
PM    & 1.0000 & 1.1253 & 0.9647 & 1.0533 & 1.3728 & 1.0145   & 1.1086 & 1.0712  & 1.0215 & 1.0718 \\
SHW   & 1.0000 & 1.1516 & 1.0803 & 1.0761 & 0.6771 & 0.5237   & 0.6120 & 0.5693  & 0.5250 & 0.5525 \\
TRV   & 1.0000 & 1.0363 & 0.9405 & 1.0220 & 0.6038 & 0.4800   & 0.5341 & 0.5131  & 0.4897 & 0.5083 \\
TSLA  & 1.0000 & 1.0401 & 1.0309 & 1.0042 & 1.0255 & 0.9596   & 0.9993 & 0.9585  & 0.9451 & 0.9513 \\
UNH   & 1.0000 & 0.9383 & 1.0250 & 0.9971 & 1.1073 & 0.9104   & 0.9712 & 0.9713  & 0.9061 & 0.9280 \\
V     & 1.0000 & 0.9524 & 0.9206 & 0.9769 & 1.0524 & 0.9350   & 0.9649 & 0.9459  & 0.9385 & 0.9505 \\
VZ    & 1.0000 & 1.0107 & 0.9412 & 0.9510 & 0.9182 & 0.7464   & 0.8242 & 0.7911  & 0.7545 & 0.7751 \\
WMT   & 1.0000 & 1.0760 & 0.9745 & 1.0707 & 1.0045 & 0.8387   & 0.8915 & 0.8597  & 0.8333 & 0.8571 \\
XOM   & 1.0000 & 1.1089 & 1.0140 & 1.0433 & 0.8792 & 0.7201   & 0.7436 & 0.7539  & 0.7149 & 0.7367 \\
\bottomrule
\end{tabular}}
\begin{scriptsize}
\parbox{\textwidth}{\emph{Note.} We forecast the daily realized variance of
each of the $40$ stocks in Table \ref{tab:stock_symbols}, with a rolling
window of $500$ observations that is moved forward one day at a time. The
sample runs from January 2, 2015, to January 30, 2026. Each entry is the
ratio of the average MSE of the model in the column to that of the HAR
model at the one-day horizon, so a value below one indicates an improvement
on the benchmark. The design otherwise follows the note to Table
\ref{mse_etf}.}
\end{scriptsize}
\end{table}

\begin{table}[H]
\centering
\caption{Out-of-sample forecast accuracy for individual stocks, QLIKE relative to the HAR model, one-day horizon.}
\label{qlike_stock_1d}
\scalebox{0.66}{
\begin{tabular}{lcccccccccc}
\toprule
  Ticker      & HAR    & HARQ   & HARJ   & HARS   & Log-AR & Log-RAR  & Log-HAR & Log-RHAR & fBm    & fOU    \\
\midrule
AAPL  & 1.0000 & 1.1742 & 0.9815 & 1.0828 & 0.9743 & 0.9014   & 0.9167 & 0.8960  & 0.8888 & 0.8919 \\
ADBE  & 1.0000 & 1.1027 & 1.0352 & 1.0286 & 0.9891 & 0.9222   & 0.9341 & 0.9186  & 0.9173 & 0.9124 \\
AMD   & 1.0000 & 1.0228 & 1.0016 & 1.0182 & 1.0478 & 0.9917   & 0.9915 & 0.9887  & 0.9856 & 0.9844 \\
AMGN  & 1.0000 & 0.9948 & 1.0647 & 0.9810 & 0.9250 & 0.8986   & 0.9114 & 0.8976  & 0.8857 & 0.8916 \\
AMZN  & 1.0000 & 0.8835 & 0.8374 & 0.9867 & 0.7908 & 0.7857   & 0.7944 & 0.7864  & 0.7857 & 0.7792 \\
AXP   & 1.0000 & 1.0827 & 0.9934 & 0.9103 & 0.7969 & 0.7521   & 0.7625 & 0.7548  & 0.7389 & 0.7430 \\
BA    & 1.0000 & 0.9763 & 1.0906 & 1.0682 & 0.8825 & 0.8034   & 0.8005 & 0.7929  & 0.7852 & 0.7885 \\
CAT   & 1.0000 & 1.1409 & 1.0190 & 1.0226 & 0.9990 & 0.9256   & 0.9426 & 0.9290  & 0.9133 & 0.9142 \\
CRM   & 1.0000 & 0.9652 & 1.0177 & 1.0260 & 0.9094 & 0.8910   & 0.9287 & 0.9133  & 0.9039 & 0.9031 \\
CSCO  & 1.0000 & 0.9979 & 1.1555 & 1.0356 & 0.9138 & 0.8722   & 0.8728 & 0.8633  & 0.8528 & 0.8606 \\
CVX   & 1.0000 & 1.0585 & 0.9555 & 0.9615 & 0.6703 & 0.6080   & 0.6320 & 0.6168  & 0.6000 & 0.6049 \\
DIS   & 1.0000 & 1.0637 & 1.0002 & 1.0383 & 0.9459 & 0.8838   & 0.8933 & 0.8835  & 0.8721 & 0.8763 \\
GE    & 1.0000 & 1.0175 & 1.0554 & 1.0747 & 1.0182 & 0.9469   & 0.9454 & 0.9370  & 0.9326 & 0.9316 \\
GOOGL & 1.0000 & 0.9552 & 0.9387 & 1.0077 & 0.9140 & 0.8863   & 0.8919 & 0.9195  & 0.8859 & 0.8865 \\
GS    & 1.0000 & 1.1898 & 1.0635 & 0.9674 & 0.7393 & 0.6925   & 0.7044 & 0.6912  & 0.6847 & 0.6862 \\
HD    & 1.0000 & 1.2561 & 0.9570 & 0.9836 & 0.7098 & 0.6596   & 0.6564 & 0.6543  & 0.6452 & 0.6470 \\
HON   & 1.0000 & 1.2326 & 1.1437 & 0.9065 & 0.8133 & 0.7841   & 0.7805 & 0.7725  & 0.7639 & 0.7733 \\
IBM   & 1.0000 & 1.0715 & 1.1200 & 1.0960 & 0.8780 & 0.8444   & 0.8445 & 0.8332  & 0.8268 & 0.8294 \\
JNJ   & 1.0000 & 1.0794 & 1.0072 & 1.0252 & 1.0198 & 0.9027   & 0.9298 & 0.9095  & 0.8831 & 0.8947 \\
JPM   & 1.0000 & 1.1344 & 1.0561 & 1.0657 & 0.8215 & 0.7656   & 0.7831 & 0.7633  & 0.7597 & 0.7621 \\
KO    & 1.0000 & 1.0062 & 1.1948 & 1.2167 & 0.8340 & 0.7537   & 0.7690 & 0.7526  & 0.7413 & 0.7485 \\
MCD   & 1.0000 & 1.0719 & 1.0707 & 0.8671 & 0.5610 & 0.5164   & 0.5289 & 0.5226  & 0.5096 & 0.5120 \\
META  & 1.0000 & 1.0253 & 0.9819 & 0.9595 & 0.9035 & 0.8448   & 0.8441 & 0.8437  & 0.8326 & 0.8343 \\
MMM   & 1.0000 & 1.0045 & 1.0178 & 1.0256 & 0.9814 & 0.9137   & 0.9213 & 0.9111  & 0.9016 & 0.9049 \\
MRK   & 1.0000 & 1.0517 & 0.9886 & 1.0188 & 0.9871 & 0.9324   & 0.9363 & 0.9283  & 0.9264 & 0.9279 \\
MSFT  & 1.0000 & 1.0066 & 0.9968 & 1.0284 & 0.9375 & 0.9157   & 0.9221 & 0.9158  & 0.9235 & 0.9093 \\
NFLX  & 1.0000 & 0.9013 & 0.8888 & 0.9726 & 0.8207 & 0.7962   & 0.8015 & 0.7905  & 0.7869 & 0.7855 \\
NKE   & 1.0000 & 1.0217 & 1.0384 & 1.1406 & 0.8228 & 0.7753   & 0.7775 & 0.7602  & 0.7523 & 0.7599 \\
NVDA  & 1.0000 & 0.9988 & 1.0186 & 1.0255 & 1.0145 & 0.9587   & 0.9665 & 0.9518  & 0.9534 & 0.9543 \\
ORCL  & 1.0000 & 1.0154 & 1.0461 & 0.9633 & 0.8833 & 0.8470   & 0.8512 & 0.8472  & 0.8455 & 0.8452 \\
PG    & 1.0000 & 0.9715 & 1.0369 & 0.9572 & 0.7058 & 0.6323   & 0.6408 & 0.6416  & 0.6215 & 0.6278 \\
PM    & 1.0000 & 0.9668 & 0.9746 & 1.0404 & 0.9533 & 0.8825   & 0.8753 & 0.8680  & 0.8475 & 0.8623 \\
SHW   & 1.0000 & 1.2064 & 0.9733 & 0.9987 & 0.7689 & 0.7310   & 0.7355 & 0.7237  & 0.7124 & 0.7164 \\
TRV   & 1.0000 & 1.1365 & 1.1319 & 0.9672 & 0.7567 & 0.7044   & 0.7104 & 0.7045  & 0.6945 & 0.6981 \\
TSLA  & 1.0000 & 1.0256 & 1.0095 & 1.0214 & 1.0456 & 0.9818   & 0.9744 & 0.9674  & 0.9555 & 0.9610 \\
UNH   & 1.0000 & 1.0469 & 1.0020 & 1.0542 & 0.9973 & 0.9004   & 0.9327 & 0.9131  & 0.8841 & 0.8933 \\
V     & 1.0000 & 1.0144 & 1.1095 & 1.0212 & 0.7601 & 0.7225   & 0.7283 & 0.7233  & 0.7107 & 0.7144 \\
VZ    & 1.0000 & 0.9263 & 0.9681 & 1.0649 & 0.7921 & 0.7302   & 0.7469 & 0.7358  & 0.7224 & 0.7253 \\
WMT   & 1.0000 & 1.0665 & 0.9878 & 1.0032 & 0.9128 & 0.8557   & 0.8528 & 0.8461  & 0.8358 & 0.8431 \\
XOM   & 1.0000 & 1.0835 & 0.9867 & 1.0004 & 0.7881 & 0.7168   & 0.7347 & 0.7268  & 0.7066 & 0.7102\\
\bottomrule
\end{tabular}}
\begin{scriptsize}
\parbox{\textwidth}{\emph{Note.} Each entry is the ratio of the average
QLIKE of the model in the column to that of the HAR model at the one-day
horizon. See the note to Table \ref{mse_stock_1d}.}
\end{scriptsize}
\end{table}

\subsection*{One-week-ahead forecasts}
\begin{table}[H]
\centering
\caption{Out-of-sample forecast accuracy for individual stocks, MSE relative to the HAR model, one-week horizon.}
\label{mse_stock_1w}
\scalebox{0.66}{
\begin{tabular}{lcccccccccc}
\toprule
  Ticker      & HAR    & HARQ   & HARJ   & HARS   & Log-AR & Log-RAR  & Log-HAR & Log-RHAR & fBm    & fOU    \\
\midrule
AAPL  & 1.0000 & 1.0021 & 1.0689 & 1.0174 & 0.8955 & 0.8142   & 0.8347 & 0.8140  & 0.8122 & 0.8077 \\
ADBE  & 1.0000 & 1.0589 & 0.9309 & 0.9707 & 0.9576 & 0.8270   & 0.8774 & 0.8513  & 0.8461 & 0.8260 \\
AMD   & 1.0000 & 0.9909 & 1.0031 & 1.0171 & 1.1200 & 1.0255   & 0.9906 & 0.9780  & 0.9856 & 0.9943 \\
AMGN  & 1.0000 & 1.1071 & 0.9453 & 1.2057 & 0.6619 & 0.6136   & 0.6312 & 0.6177  & 0.5954 & 0.6114 \\
AMZN  & 1.0000 & 0.9920 & 0.9903 & 1.0185 & 0.9819 & 0.9657   & 0.9622 & 0.9622  & 0.9631 & 0.9608 \\
AXP   & 1.0000 & 1.2272 & 1.0219 & 1.1261 & 0.9556 & 0.7770   & 0.8784 & 0.8177  & 0.7580 & 0.7771 \\
BA    & 1.0000 & 0.7964 & 0.7920 & 0.9627 & 0.6679 & 0.5614   & 0.5918 & 0.5694  & 0.5343 & 0.5585 \\
CAT   & 1.0000 & 0.9570 & 1.0872 & 1.0379 & 0.9900 & 0.7957   & 0.8652 & 0.8251  & 0.8283 & 0.8238 \\
CRM   & 1.0000 & 0.9750 & 1.0065 & 1.0064 & 0.9300 & 0.8647   & 0.8867 & 0.8751  & 0.8680 & 0.8590 \\
CSCO  & 1.0000 & 0.9462 & 1.0520 & 1.0720 & 0.9462 & 0.8105   & 0.8545 & 0.8193  & 0.7881 & 0.8192 \\
CVX   & 1.0000 & 1.0251 & 1.0246 & 0.8796 & 1.1650 & 0.8179   & 0.9382 & 0.8800  & 0.8090 & 0.8526 \\
DIS   & 1.0000 & 1.0478 & 0.9594 & 1.1034 & 1.0064 & 0.8079   & 0.8275 & 0.7840  & 0.7480 & 0.8212 \\
GE    & 1.0000 & 1.0081 & 1.0034 & 1.0351 & 1.0351 & 0.8512   & 0.9326 & 0.8875  & 0.8438 & 0.8445 \\
GOOGL & 1.0000 & 0.9858 & 0.9766 & 1.0100 & 1.0030 & 0.9472   & 0.9703 & 0.9620  & 0.9486 & 0.9454 \\
GS    & 1.0000 & 0.9593 & 0.9948 & 1.2138 & 0.8736 & 0.7561   & 0.8410 & 0.7901  & 0.7555 & 0.7656 \\
HD    & 1.0000 & 1.0544 & 0.9386 & 1.1288 & 0.9091 & 0.7607   & 0.8435 & 0.7963  & 0.7464 & 0.7692 \\
HON   & 1.0000 & 1.0352 & 0.9735 & 0.8476 & 0.6499 & 0.5554   & 0.6027 & 0.5803  & 0.5320 & 0.5558 \\
IBM   & 1.0000 & 0.9664 & 0.9617 & 1.0453 & 0.9220 & 0.7370   & 0.7996 & 0.7721  & 0.7227 & 0.7504 \\
JNJ   & 1.0000 & 1.1114 & 1.0098 & 1.1366 & 1.0282 & 0.8410   & 0.8781 & 0.8435  & 0.8324 & 0.8677 \\
JPM   & 1.0000 & 0.9624 & 0.9927 & 1.0360 & 0.9894 & 0.7983   & 0.8428 & 0.8100  & 0.7867 & 0.8208 \\
KO    & 1.0000 & 0.9992 & 0.9540 & 1.0961 & 0.9228 & 0.7451   & 0.8052 & 0.7298  & 0.7248 & 0.7583 \\
MCD   & 1.0000 & 1.1642 & 1.2090 & 1.2028 & 0.8301 & 0.7486   & 0.8446 & 0.8398  & 0.7035 & 0.7354 \\
META  & 1.0000 & 1.0349 & 1.0093 & 1.0065 & 0.9630 & 0.8711   & 0.8813 & 0.8707  & 0.8692 & 0.8556 \\
MMM   & 1.0000 & 1.0720 & 0.9875 & 1.0869 & 1.0720 & 0.9233   & 0.9820 & 0.9518  & 0.9138 & 0.9369 \\
MRK   & 1.0000 & 1.0670 & 1.0549 & 1.0567 & 0.9810 & 0.8183   & 0.8704 & 0.8317  & 0.7968 & 0.8311 \\
MSFT  & 1.0000 & 1.0482 & 1.0174 & 1.0313 & 0.9044 & 0.8255   & 0.8660 & 0.8491  & 0.8183 & 0.8152 \\
NFLX  & 1.0000 & 0.9903 & 0.9878 & 1.0562 & 0.9976 & 0.9757   & 0.9740 & 0.9711  & 0.9705 & 0.9671 \\
NKE   & 1.0000 & 1.0345 & 1.2748 & 1.2223 & 1.0139 & 0.8744   & 0.9212 & 0.8960  & 0.8560 & 0.8809 \\
NVDA  & 1.0000 & 0.9665 & 0.9945 & 1.0184 & 0.9947 & 0.9046   & 0.9053 & 0.8980  & 0.9131 & 0.8934 \\
ORCL  & 1.0000 & 0.9724 & 0.9620 & 1.0233 & 0.9806 & 0.8796   & 0.8873 & 0.8715  & 0.8839 & 0.8842 \\
PG    & 1.0000 & 1.0113 & 1.0454 & 1.1225 & 0.9366 & 0.7776   & 0.8460 & 0.7906  & 0.7387 & 0.7724 \\
PM    & 1.0000 & 0.9303 & 0.9624 & 1.0671 & 0.8252 & 0.6794   & 0.7408 & 0.7072  & 0.6682 & 0.6948 \\
SHW   & 1.0000 & 1.0425 & 0.8833 & 0.7822 & 0.7920 & 0.6704   & 0.7704 & 0.7166  & 0.6518 & 0.6664 \\
TRV   & 1.0000 & 0.9497 & 0.9621 & 0.8535 & 0.8321 & 0.6910   & 0.7678 & 0.7228  & 0.6749 & 0.6932 \\
TSLA  & 1.0000 & 0.9751 & 1.0082 & 1.0170 & 0.9468 & 0.8816   & 0.8878 & 0.8817  & 0.8793 & 0.8694 \\
UNH   & 1.0000 & 0.9552 & 0.9596 & 1.0091 & 0.9818 & 0.7542   & 0.8087 & 0.8065  & 0.7245 & 0.7541 \\
V     & 1.0000 & 0.9644 & 1.0707 & 0.8532 & 0.6556 & 0.6076   & 0.6229 & 0.6149  & 0.5995 & 0.6095 \\
VZ    & 1.0000 & 1.0158 & 1.0315 & 0.9990 & 0.8918 & 0.7495   & 0.8275 & 0.7882  & 0.7440 & 0.7565 \\
WMT   & 1.0000 & 1.0113 & 0.9939 & 1.0241 & 0.8876 & 0.7831   & 0.8309 & 0.7868  & 0.7591 & 0.7755 \\
XOM   & 1.0000 & 1.1268 & 1.1022 & 1.1077 & 1.0298 & 0.7528   & 0.8711 & 0.8281  & 0.7191 & 0.7662\\
\bottomrule
\end{tabular}}
\begin{scriptsize}
\parbox{\textwidth}{\emph{Note.} Each entry is the ratio of the average
MSE of the model in the column to that of the HAR model at the one-week
horizon. See the note to Table \ref{mse_stock_1d}.}
\end{scriptsize}
\end{table}

\begin{table}[H]
\centering
\caption{Out-of-sample forecast accuracy for individual stocks, QLIKE relative to the HAR model, one-week horizon.}
\label{qlike_stock_1w}
\scalebox{0.66}{
\begin{tabular}{lcccccccccc}
\toprule
  Ticker      & HAR    & HARQ   & HARJ   & HARS   & Log-AR & Log-RAR  & Log-HAR & Log-RHAR & fBm    & fOU    \\
\midrule
AAPL  & 1.0000 & 0.9967 & 1.0060 & 1.0493 & 1.1823 & 0.9609   & 0.9364 & 0.9064  & 0.8883 & 0.9281 \\
ADBE  & 1.0000 & 1.0457 & 1.0000 & 1.0111 & 1.1304 & 0.9340   & 0.9459 & 0.9174  & 0.9029 & 0.9059 \\
AMD   & 1.0000 & 0.9951 & 0.9992 & 1.0135 & 1.1769 & 1.0455   & 1.0015 & 0.9947  & 1.0119 & 1.0146 \\
AMGN  & 1.0000 & 0.9682 & 1.0136 & 1.0216 & 1.1170 & 0.9014   & 0.9179 & 0.8671  & 0.8462 & 0.8869 \\
AMZN  & 1.0000 & 0.9652 & 0.9429 & 0.9946 & 1.0127 & 0.8760   & 0.8914 & 0.8715  & 0.8400 & 0.8545 \\
AXP   & 1.0000 & 1.0149 & 1.0291 & 1.0132 & 1.1785 & 0.8488   & 0.8424 & 0.8098  & 0.7612 & 0.8100 \\
BA    & 1.0000 & 0.9540 & 0.9971 & 1.0386 & 1.2569 & 0.8349   & 0.8566 & 0.7954  & 0.7099 & 0.7599 \\
CAT   & 1.0000 & 0.9776 & 0.9788 & 0.9901 & 1.1648 & 0.8657   & 0.9210 & 0.8868  & 0.8538 & 0.8573 \\
CRM   & 1.0000 & 1.0155 & 1.0030 & 1.0118 & 1.0847 & 1.0006   & 0.9923 & 0.9921  & 0.9785 & 0.9959 \\
CSCO  & 1.0000 & 0.9848 & 1.0244 & 1.0192 & 1.1706 & 0.9320   & 0.9078 & 0.8862  & 0.8384 & 0.8949 \\
CVX   & 1.0000 & 0.9690 & 1.0111 & 0.9651 & 0.9338 & 0.6511   & 0.7304 & 0.6803  & 0.6363 & 0.6475 \\
DIS   & 1.0000 & 1.0058 & 0.9547 & 0.9535 & 1.0735 & 0.7995   & 0.7988 & 0.7633  & 0.7336 & 0.7875 \\
GE    & 1.0000 & 1.0070 & 0.9946 & 1.0120 & 1.2612 & 0.9720   & 0.9538 & 0.9195  & 0.9085 & 0.9193 \\
GOOGL & 1.0000 & 0.9945 & 0.9837 & 1.0119 & 1.0454 & 0.9010   & 0.9403 & 0.9442  & 0.8772 & 0.8929 \\
GS    & 1.0000 & 1.0765 & 0.9965 & 1.0388 & 1.1493 & 0.9184   & 0.9176 & 0.8864  & 0.8374 & 0.8842 \\
HD    & 1.0000 & 1.0980 & 1.0348 & 1.1283 & 1.1398 & 0.8285   & 0.8202 & 0.7799  & 0.7561 & 0.7839 \\
HON   & 1.0000 & 1.0392 & 0.9947 & 0.9742 & 1.1325 & 0.8596   & 0.8468 & 0.8241  & 0.7535 & 0.8091 \\
IBM   & 1.0000 & 1.0362 & 0.9860 & 1.0279 & 1.1678 & 0.8845   & 0.8437 & 0.8343  & 0.7963 & 0.8278 \\
JNJ   & 1.0000 & 1.0078 & 1.0099 & 1.0171 & 1.2609 & 0.8743   & 0.9067 & 0.8886  & 0.8296 & 0.8669 \\
JPM   & 1.0000 & 1.0655 & 1.0020 & 1.0376 & 1.0913 & 0.8283   & 0.8242 & 0.7969  & 0.7679 & 0.8023 \\
KO    & 1.0000 & 1.0539 & 1.0164 & 1.0520 & 1.1746 & 0.7726   & 0.7972 & 0.7671  & 0.7169 & 0.7570 \\
MCD   & 1.0000 & 1.0084 & 1.0492 & 0.9821 & 1.0066 & 0.7216   & 0.7484 & 0.7023  & 0.6474 & 0.6815 \\
META  & 1.0000 & 0.9916 & 0.9917 & 0.9941 & 1.1020 & 0.8953   & 0.8844 & 0.8668  & 0.8226 & 0.8605 \\
MMM   & 1.0000 & 0.9882 & 0.9884 & 1.0238 & 1.1823 & 0.9095   & 0.9245 & 0.8915  & 0.8311 & 0.8832 \\
MRK   & 1.0000 & 1.0385 & 1.0083 & 1.0383 & 1.2419 & 0.9377   & 0.9319 & 0.8925  & 0.8857 & 0.9254 \\
MSFT  & 1.0000 & 1.0331 & 0.9618 & 1.0052 & 1.0559 & 0.8950   & 0.9296 & 0.8979  & 0.8561 & 0.8613 \\
NFLX  & 1.0000 & 0.9736 & 0.9575 & 1.0018 & 1.0459 & 0.8818   & 0.8765 & 0.8616  & 0.8396 & 0.8429 \\
NKE   & 1.0000 & 1.0050 & 1.0089 & 1.0574 & 1.1330 & 0.8769   & 0.8377 & 0.8148  & 0.7767 & 0.8270 \\
NVDA  & 1.0000 & 0.9878 & 1.0025 & 1.0181 & 1.1274 & 0.9601   & 0.9418 & 0.9269  & 0.9387 & 0.9432 \\
ORCL  & 1.0000 & 0.9583 & 0.9749 & 0.9904 & 1.0560 & 0.8861   & 0.8889 & 0.8687  & 0.8339 & 0.8796 \\
PG    & 1.0000 & 1.0878 & 0.9987 & 1.0644 & 1.2241 & 0.7400   & 0.7646 & 0.7364  & 0.6885 & 0.7391 \\
PM    & 1.0000 & 0.9152 & 1.0233 & 1.0523 & 1.2454 & 0.8383   & 0.8558 & 0.8049  & 0.7511 & 0.8040 \\
SHW   & 1.0000 & 1.0707 & 0.9940 & 0.9705 & 1.1731 & 0.8652   & 0.8934 & 0.8518  & 0.7683 & 0.8116 \\
TRV   & 1.0000 & 1.0488 & 1.0012 & 0.9573 & 1.1030 & 0.8013   & 0.8221 & 0.7797  & 0.7469 & 0.7757 \\
TSLA  & 1.0000 & 0.9781 & 1.0034 & 1.0104 & 1.2122 & 1.0100   & 0.9600 & 0.9529  & 0.9127 & 0.9465 \\
UNH   & 1.0000 & 1.0098 & 1.0639 & 1.0209 & 1.1339 & 0.8895   & 0.8710 & 0.8543  & 0.8106 & 0.8476 \\
V     & 1.0000 & 1.0488 & 1.0003 & 1.0085 & 1.0214 & 0.8024   & 0.8007 & 0.7834  & 0.7397 & 0.7796 \\
VZ    & 1.0000 & 0.9709 & 1.0041 & 1.0494 & 1.0523 & 0.7990   & 0.8704 & 0.8065  & 0.7586 & 0.7797 \\
WMT   & 1.0000 & 1.0329 & 1.0255 & 1.0051 & 1.2394 & 0.9349   & 0.8880 & 0.8793  & 0.8351 & 0.8794 \\
XOM   & 1.0000 & 1.1703 & 1.0577 & 1.0362 & 0.9822 & 0.6765   & 0.7382 & 0.6906  & 0.6401 & 0.6582\\
\bottomrule
\end{tabular}}
\begin{scriptsize}
\parbox{\textwidth}{\emph{Note.} Each entry is the ratio of the average
QLIKE of the model in the column to that of the HAR model at the one-week
horizon. See the note to Table \ref{mse_stock_1d}.}
\end{scriptsize}
\end{table}

\subsection*{One-month-ahead forecasts}
\begin{table}[H]
\centering
\caption{Out-of-sample forecast accuracy for individual stocks, MSE relative to the HAR model, one-month horizon.}
\label{mse_stock_1m}
\scalebox{0.66}{
\begin{tabular}{lcccccccccc}
\toprule
  Ticker      & HAR    & HARQ   & HARJ   & HARS   & Log-AR & Log-RAR  & Log-HAR & Log-RHAR & fBm    & fOU    \\
\midrule
AAPL  & 1.0000 & 0.9283 & 1.0480 & 0.9823 & 0.7845 & 0.7729   & 0.7733 & 0.7794  & 0.8239 & 0.7657 \\
ADBE  & 1.0000 & 1.0972 & 0.9966 & 1.0077 & 0.9870 & 0.9750   & 0.9813 & 0.9925  & 1.0581 & 0.9650 \\
AMD   & 1.0000 & 1.0028 & 0.9972 & 1.0099 & 1.0210 & 1.0071   & 0.9757 & 0.9534  & 0.9972 & 0.9898 \\
AMGN  & 1.0000 & 1.3088 & 1.0038 & 0.9975 & 0.9490 & 0.9466   & 0.9449 & 0.9716  & 0.9967 & 0.9453 \\
AMZN  & 1.0000 & 1.0011 & 1.0000 & 1.0021 & 0.9910 & 0.9857   & 0.9819 & 0.9825  & 0.9852 & 0.9798 \\
AXP   & 1.0000 & 0.9816 & 1.0073 & 1.0241 & 0.8707 & 0.8769   & 0.8630 & 0.9275  & 0.8801 & 0.8572 \\
BA    & 1.0000 & 0.9360 & 1.0763 & 1.0690 & 0.3338 & 0.3280   & 0.3288 & 0.3458  & 0.3389 & 0.3254 \\
CAT   & 1.0000 & 0.9860 & 0.9990 & 1.0120 & 0.9415 & 0.9512   & 0.9414 & 0.9602  & 0.9729 & 0.9378 \\
CRM   & 1.0000 & 1.0628 & 1.0168 & 1.0133 & 0.9627 & 0.9525   & 0.9559 & 0.9830  & 1.0001 & 0.9472 \\
CSCO  & 1.0000 & 0.9858 & 0.9996 & 1.0012 & 0.9238 & 0.9169   & 0.9211 & 0.9238  & 0.9988 & 0.9156 \\
CVX   & 1.0000 & 1.0964 & 1.0098 & 1.0116 & 0.9984 & 1.0094   & 0.9644 & 1.1979  & 1.0394 & 0.9851 \\
DIS   & 1.0000 & 0.9734 & 0.9354 & 1.0207 & 0.5676 & 0.5573   & 0.5594 & 0.5554  & 0.5817 & 0.5539 \\
GE    & 1.0000 & 0.9826 & 1.0066 & 1.0118 & 0.9910 & 0.9384   & 0.9421 & 1.0142  & 0.9916 & 0.9272 \\
GOOGL & 1.0000 & 0.9980 & 1.0013 & 1.0034 & 0.9858 & 0.9799   & 0.9875 & 0.9895  & 1.0166 & 0.9807 \\
GS    & 1.0000 & 0.8621 & 0.9797 & 0.9782 & 0.6401 & 0.6366   & 0.6442 & 0.6644  & 0.6714 & 0.6407 \\
HD    & 1.0000 & 1.0548 & 0.9917 & 1.0021 & 0.9515 & 0.9676   & 0.9460 & 1.0284  & 1.0115 & 0.9536 \\
HON   & 1.0000 & 1.0003 & 0.9884 & 1.0192 & 0.8084 & 0.7992   & 0.7983 & 0.8627  & 0.8541 & 0.8001 \\
IBM   & 1.0000 & 0.8878 & 0.9847 & 0.9882 & 0.7201 & 0.7126   & 0.7143 & 0.7643  & 0.7513 & 0.7110 \\
JNJ   & 1.0000 & 1.0563 & 1.0002 & 1.0021 & 0.9714 & 0.9736   & 0.9745 & 0.9719  & 1.0238 & 0.9747 \\
JPM   & 1.0000 & 0.9350 & 0.9975 & 0.8886 & 0.7737 & 0.7651   & 0.7663 & 0.7705  & 0.7944 & 0.7613 \\
KO    & 1.0000 & 1.1218 & 1.0032 & 1.0005 & 0.9816 & 1.0272   & 0.9772 & 0.9908  & 1.0196 & 0.9821 \\
MCD   & 1.0000 & 1.0364 & 0.9981 & 0.9957 & 0.9761 & 0.9779   & 0.9729 & 1.1912  & 1.0587 & 0.9836 \\
META  & 1.0000 & 0.9783 & 0.9973 & 1.0125 & 0.9725 & 0.9414   & 0.9350 & 0.9250  & 0.9691 & 0.9113 \\
MMM   & 1.0000 & 0.9966 & 1.0058 & 1.0088 & 0.9536 & 0.9494   & 0.9464 & 0.9844  & 0.9774 & 0.9511 \\
MRK   & 1.0000 & 1.0043 & 1.0039 & 1.0017 & 0.9956 & 0.9904   & 0.9789 & 1.0242  & 1.0092 & 0.9885 \\
MSFT  & 1.0000 & 0.9686 & 1.0008 & 1.0156 & 0.9509 & 0.9445   & 0.9517 & 1.0032  & 1.0429 & 0.9386 \\
NFLX  & 1.0000 & 1.0008 & 1.0058 & 1.0769 & 1.0015 & 0.9902   & 0.9815 & 0.9784  & 0.9771 & 0.9741 \\
NKE   & 1.0000 & 0.9630 & 1.0381 & 1.0242 & 0.7951 & 0.8038   & 0.7960 & 0.8587  & 0.8296 & 0.7970 \\
NVDA  & 1.0000 & 0.9892 & 1.0097 & 1.0099 & 0.9956 & 0.9749   & 0.9636 & 0.9655  & 0.9841 & 0.9532 \\
ORCL  & 1.0000 & 0.9844 & 1.0046 & 1.0044 & 0.9461 & 0.9490   & 0.9374 & 0.9364  & 1.0000 & 0.9507 \\
PG    & 1.0000 & 1.0577 & 1.0026 & 1.0016 & 0.9755 & 1.0194   & 0.9673 & 1.0045  & 1.0085 & 0.9810 \\
PM    & 1.0000 & 1.0069 & 0.9948 & 1.0027 & 0.9709 & 0.9772   & 0.9626 & 1.1284  & 0.9921 & 0.9635 \\
SHW   & 1.0000 & 0.9997 & 0.9975 & 1.0011 & 0.9853 & 0.9774   & 0.9635 & 1.0960  & 1.0260 & 0.9738 \\
TRV   & 1.0000 & 0.9986 & 0.9980 & 1.0083 & 0.9805 & 0.9611   & 0.9597 & 0.9982  & 0.9890 & 0.9481 \\
TSLA  & 1.0000 & 0.9975 & 1.0047 & 1.0081 & 0.6862 & 0.6798   & 0.6888 & 0.6988  & 0.7257 & 0.6742 \\
UNH   & 1.0000 & 0.9660 & 0.9964 & 0.9961 & 0.8612 & 0.8478   & 0.8484 & 1.0076  & 0.8870 & 0.8522 \\
V     & 1.0000 & 1.2072 & 1.0147 & 1.0017 & 0.9752 & 0.9731   & 0.9695 & 0.9880  & 1.0069 & 0.9721 \\
VZ    & 1.0000 & 0.9545 & 1.0017 & 1.0127 & 0.9293 & 0.9307   & 0.9365 & 1.0092  & 0.9805 & 0.9301 \\
WMT   & 1.0000 & 0.9101 & 1.0194 & 0.9847 & 0.8165 & 0.8149   & 0.8190 & 0.8211  & 0.8363 & 0.8073 \\
XOM   & 1.0000 & 0.9886 & 1.0603 & 1.0215 & 0.9780 & 0.9492   & 0.9298 & 1.2371  & 0.9721 & 0.9383\\
\bottomrule
\end{tabular}}
\begin{scriptsize}
\parbox{\textwidth}{\emph{Note.} Each entry is the ratio of the average
MSE of the model in the column to that of the HAR model at the one-month
horizon. See the note to Table \ref{mse_stock_1d}.}
\end{scriptsize}
\end{table}

\begin{table}[H]
\centering
\caption{Out-of-sample forecast accuracy for individual stocks, QLIKE relative to the HAR model, one-month horizon.}
\label{qlike_stock_1m}
\scalebox{0.66}{
\begin{tabular}{lcccccccccc}
\toprule
  Ticker      & HAR    & HARQ   & HARJ   & HARS   & Log-AR & Log-RAR  & Log-HAR & Log-RHAR & fBm    & fOU    \\
\midrule
AAPL  & 1.0000 & 1.0040 & 1.0147 & 1.0110 & 1.0589 & 1.0045   & 0.9340 & 0.9091  & 0.8709 & 0.9393 \\
ADBE  & 1.0000 & 1.0001 & 1.0243 & 1.0154 & 1.0351 & 0.9933   & 0.9932 & 0.9781  & 0.9828 & 0.9560 \\
AMD   & 1.0000 & 1.0017 & 0.9962 & 1.0099 & 1.0353 & 1.0171   & 1.0037 & 0.9516  & 0.9712 & 0.9984 \\
AMGN  & 1.0000 & 1.0074 & 1.0067 & 1.0102 & 0.9521 & 0.9337   & 0.9434 & 0.9251  & 0.9264 & 0.9055 \\
AMZN  & 1.0000 & 1.0255 & 1.0209 & 1.0054 & 0.9892 & 0.9462   & 0.9506 & 0.9263  & 0.8498 & 0.9084 \\
AXP   & 1.0000 & 1.0009 & 0.9946 & 1.0159 & 1.0234 & 0.9500   & 0.9655 & 0.9456  & 0.8846 & 0.9092 \\
BA    & 1.0000 & 0.9902 & 1.0001 & 1.0119 & 1.0639 & 0.9742   & 0.9838 & 0.9386  & 0.8528 & 0.8660 \\
CAT   & 1.0000 & 1.0158 & 1.0178 & 1.0221 & 0.9617 & 0.9388   & 0.9772 & 0.9664  & 0.9623 & 0.9355 \\
CRM   & 1.0000 & 1.0215 & 1.0252 & 1.0166 & 1.0399 & 1.0038   & 0.9823 & 0.9939  & 0.9357 & 0.9717 \\
CSCO  & 1.0000 & 0.9881 & 0.9925 & 0.9961 & 0.8677 & 0.8315   & 0.8386 & 0.7997  & 0.7659 & 0.7784 \\
CVX   & 1.0000 & 1.0042 & 1.0103 & 1.0117 & 1.0804 & 0.9674   & 0.9471 & 0.9629  & 0.9137 & 0.8906 \\
DIS   & 1.0000 & 1.0252 & 1.0196 & 1.0222 & 1.0168 & 0.9553   & 0.9600 & 0.9253  & 0.8244 & 0.9042 \\
GE    & 1.0000 & 0.9998 & 0.9951 & 1.0079 & 1.2178 & 0.9810   & 0.9437 & 0.9234  & 0.8797 & 0.8991 \\
GOOGL & 1.0000 & 0.9893 & 1.0049 & 1.0064 & 0.9574 & 0.9277   & 0.9695 & 0.9525  & 0.9089 & 0.9186 \\
GS    & 1.0000 & 1.0225 & 0.9962 & 0.9951 & 0.9347 & 0.9128   & 0.9766 & 0.9531  & 0.9232 & 0.9164 \\
HD    & 1.0000 & 0.9951 & 0.9950 & 1.0220 & 0.9339 & 0.8947   & 0.9173 & 0.8938  & 0.8812 & 0.8415 \\
HON   & 1.0000 & 1.0366 & 0.9887 & 1.0035 & 1.0013 & 0.9468   & 0.9369 & 0.9126  & 0.8372 & 0.8875 \\
IBM   & 1.0000 & 1.0058 & 0.9975 & 1.0134 & 1.0654 & 1.0012   & 0.9561 & 0.9148  & 0.8497 & 0.9184 \\
JNJ   & 1.0000 & 0.9900 & 0.9921 & 1.0293 & 0.9256 & 0.9150   & 0.9425 & 0.9356  & 0.9502 & 0.8862 \\
JPM   & 1.0000 & 0.9922 & 1.0001 & 0.9972 & 0.9453 & 0.8956   & 0.9286 & 0.9179  & 0.9226 & 0.8561 \\
KO    & 1.0000 & 1.0026 & 1.0081 & 1.0001 & 0.9590 & 0.9139   & 0.9530 & 0.9351  & 0.9392 & 0.8746 \\
MCD   & 1.0000 & 1.0094 & 1.0010 & 1.0040 & 0.9271 & 0.8940   & 0.9290 & 0.9290  & 0.9034 & 0.8189 \\
META  & 1.0000 & 1.0139 & 1.0061 & 1.0140 & 0.9715 & 0.9104   & 0.9017 & 0.8567  & 0.8254 & 0.8455 \\
MMM   & 1.0000 & 0.9904 & 0.9972 & 1.0110 & 1.0611 & 0.9980   & 0.9549 & 0.9424  & 0.8591 & 0.9427 \\
MRK   & 1.0000 & 1.0024 & 0.9970 & 0.9993 & 1.0144 & 0.9802   & 0.9553 & 0.9388  & 0.9283 & 0.9489 \\
MSFT  & 1.0000 & 0.9830 & 0.9794 & 0.9736 & 0.9309 & 0.8917   & 0.9025 & 0.8658  & 0.8448 & 0.8201 \\
NFLX  & 1.0000 & 1.0014 & 0.9996 & 1.0083 & 1.0679 & 0.9907   & 0.9059 & 0.8929  & 0.8444 & 0.8740 \\
NKE   & 1.0000 & 1.0052 & 1.0213 & 1.0155 & 1.0624 & 1.0284   & 0.9887 & 0.9818  & 0.9395 & 0.9710 \\
NVDA  & 1.0000 & 0.9860 & 0.9988 & 1.0160 & 1.0437 & 0.9983   & 1.0073 & 0.9616  & 0.9559 & 0.9706 \\
ORCL  & 1.0000 & 0.9994 & 0.9959 & 1.0135 & 0.9723 & 0.9778   & 0.9972 & 0.9840  & 0.9896 & 0.9963 \\
PG    & 1.0000 & 1.0278 & 1.0097 & 1.0092 & 0.9784 & 0.9100   & 0.9162 & 0.8981  & 0.8700 & 0.8607 \\
PM    & 1.0000 & 0.9865 & 0.9970 & 1.0104 & 1.0011 & 0.9326   & 0.9219 & 0.8957  & 0.8671 & 0.8724 \\
SHW   & 1.0000 & 0.9989 & 0.9959 & 1.0028 & 0.9603 & 0.9173   & 0.9203 & 0.9267  & 0.8934 & 0.8611 \\
TRV   & 1.0000 & 1.0097 & 1.0083 & 1.0117 & 1.0080 & 0.9298   & 0.9459 & 0.9312  & 0.9196 & 0.8756 \\
TSLA  & 1.0000 & 0.9967 & 1.0291 & 1.0092 & 0.9834 & 0.9493   & 0.9155 & 0.8627  & 0.8230 & 0.8906 \\
UNH   & 1.0000 & 0.9876 & 1.0058 & 1.0495 & 1.0702 & 0.9852   & 0.8898 & 0.8697  & 0.7816 & 0.8954 \\
V     & 1.0000 & 0.9913 & 1.0172 & 0.9931 & 0.9347 & 0.8766   & 0.8696 & 0.8543  & 0.8077 & 0.8244 \\
VZ    & 1.0000 & 0.9940 & 1.0031 & 1.0093 & 0.8879 & 0.8510   & 0.8996 & 0.8777  & 0.8688 & 0.8239 \\
WMT   & 1.0000 & 1.0053 & 1.0222 & 1.0107 & 0.9340 & 0.8989   & 0.9406 & 0.8693  & 0.8474 & 0.8353 \\
XOM   & 1.0000 & 1.0880 & 1.0361 & 1.0401 & 1.1653 & 1.0080   & 0.9510 & 0.9371  & 0.8668 & 0.8983\\
\bottomrule
\end{tabular}}
\begin{scriptsize}
\parbox{\textwidth}{\emph{Note.} Each entry is the ratio of the average
QLIKE of the model in the column to that of the HAR model at the one-month
horizon. See the note to Table \ref{mse_stock_1d}.}
\end{scriptsize}
\end{table}

\section{Additional simulation results}
\label{sec:app:sim}

This appendix collects additional results for the simulation studies in Section \ref{sec:simulation}. Table \ref{tab:sim_theta_app} reports the parameter recovery results for the fOU process with $\kappa = 5$. 

\begin{table}[H]
\centering
\caption{Recovering the Hurst parameter: fOU with $\kappa = 5$.}
\label{tab:sim_theta_app}
\scalebox{0.78}{
\begin{tabular}{cccccccccc}
\toprule
 & & \multicolumn{4}{c}{$n=500$} & \multicolumn{4}{c}{$n=4{,}000$} \\
\cmidrule(lr){3-6} \cmidrule(lr){7-10}
 & & \multicolumn{2}{c}{RAR} & \multicolumn{2}{c}{RHAR} & \multicolumn{2}{c}{RAR} & \multicolumn{2}{c}{RHAR} \\
\cmidrule(lr){3-4} \cmidrule(lr){5-6} \cmidrule(lr){7-8} \cmidrule(lr){9-10}
$H$ & $c_1(H)$ & $\hat\theta$ & $H_{\mathrm{imp}}$ & $\hat\theta$ & $H_{\mathrm{imp}}$ & $\hat\theta$ & $H_{\mathrm{imp}}$ & $\hat\theta$ & $H_{\mathrm{imp}}$ \\
\midrule
\multicolumn{10}{c}{fOU with $\kappa=5$} \\
\midrule
0.1 & -0.657 & -0.579 (0.074) & 0.138 & -0.403 (0.254) & 0.241 & -0.594 (0.024) & 0.129 & -0.422 (0.085) & 0.223 \\
0.2 & -0.459 & -0.406 (0.064) & 0.232 & -0.268 (0.137) & 0.319 & -0.414 (0.020) & 0.226 & -0.268 (0.046) & 0.316 \\
0.3 & -0.293 & -0.257 (0.056) & 0.323 & -0.160 (0.099) & 0.388 & -0.266 (0.019) & 0.317 & -0.172 (0.033) & 0.380 \\
0.4 & -0.142 & -0.125 (0.048) & 0.412 & -0.083 (0.080) & 0.437 & -0.129 (0.018) & 0.409 & -0.081 (0.027) & 0.442 \\
\bottomrule
\end{tabular}}
\begin{scriptsize}
\parbox{\textwidth}{\emph{Note.} We simulate 500 replications per cell of the fOU process with $\kappa = 5$ at the daily frequency ($\Delta=1/250$) and estimate the ``rough'' AR (RAR) and ``rough'' HAR (RHAR) models by exact maximum likelihood via the Kalman filter. The table shows the Monte Carlo average of $\hat\theta$ (standard deviation in parentheses) next to the first Wold coefficient $c_1(H)$ of fGn, and the implied Hurst parameter $H_{\mathrm{imp}}$, computed by inverting the $c_1(\cdot)$ mapping at each replication's $\hat\theta$ and averaging, with the inversion restricted to $H \in (0, 1/2)$. Replications in which the likelihood optimization fails to converge, at most six of $500$ per cell, are excluded.}
\end{scriptsize}
\end{table}

\section{Recovering Wold coefficients from the spectral density via Szegö factorization}
\label{app:szego_factor}

This appendix summarizes how the Wold (innovations) coefficients in
\eqref{eq:wold} can be computed from the spectral density $f$ of a purely
nondeterministic stationary process. The exposition follows standard results
on Szegö factorization, see, e.g., \citet{bingham:2012a}.

\subsection{Szegö's condition and the outer (Szegö) function}

Let $f(\lambda)$ be the spectral density on $[-\pi,\pi]$ of a stationary process
$\{X_t\}$. Assume Szegö's condition
\begin{equation}
\log f \in L^1([-\pi,\pi]).
\label{eq:szego_condition}
\end{equation}
Define the Szegö (outer) function
\begin{equation}
h(z)
:= \exp\!\left(\frac{1}{4\pi}\int_{-\pi}^{\pi}
\frac{e^{i\lambda}+z}{e^{i\lambda}-z}\,\log f(\lambda)\,d\lambda\right),
\qquad z\in \mathbb{D},
\label{eq:szego_function}
\end{equation}
where $\mathbb{D}=\{z\in\mathbb{C}:|z|<1\}$.
Under \eqref{eq:szego_condition}, the radial boundary values exist for a.e.
$\lambda$, i.e.,
\begin{equation*}
h(e^{i\lambda}) := \lim_{r\uparrow 1} h(re^{i\lambda})
\end{equation*}
exists for a.e.\ $\lambda$, and satisfy the boundary modulus identity
\begin{equation}
|h(e^{i\lambda})|^2 = f(\lambda)
\qquad \text{for a.e.\ }\lambda\in[-\pi,\pi].
\label{eq:boundary_modulus}
\end{equation}

Moreover, $h$ admits a power-series expansion
\begin{equation*}
h(z) = h(0)\sum_{k=0}^{\infty} c_k z^k,
\qquad c_0=1,
\end{equation*}
and the coefficients $\{c_k\}$ coincide with the normalized Wold coefficients
of the innovations representation.

\subsection{Normalization and the innovation variance}

Let
\begin{equation}
a_k := \int_{-\pi}^{\pi} \log f(\lambda)\,e^{-ik\lambda}\,\frac{d\lambda}{2\pi}
\qquad (k\in\mathbb{Z})
\label{eq:cepstral}
\end{equation}
denote the Fourier (cepstral) coefficients of $\log f$.
Evaluating \eqref{eq:szego_function} at $z=0$ gives
\begin{equation*}
h(0)
= \exp\!\left(\frac{1}{4\pi}\int_{-\pi}^{\pi}\log f(\lambda)\,d\lambda\right)
= \exp\!\left(\frac{a_0}{2}\right),
\qquad |h(0)|^2 = e^{a_0}.
\end{equation*}

Define the normalized spectral factor
\begin{equation}
C(z):=\frac{h(z)}{h(0)}=\sum_{k=0}^{\infty}c_k z^k,
\qquad C(0)=c_0=1.
\label{eq:normalized_factor}
\end{equation}
Then, using \eqref{eq:boundary_modulus} and \eqref{eq:normalized_factor},
\begin{equation}
f(\lambda)=|h(e^{i\lambda})|^2 = |h(0)|^2\,|C(e^{i\lambda})|^2
\qquad \text{a.e.}
\label{eq:f_factorized_basic}
\end{equation}
On the other hand, for the Wold representation
$X_t=\sum_{j\ge 0} c_j\varepsilon_{t-j}$ with $\operatorname{Var}(\varepsilon_t)=\sigma_\varepsilon^2$,
the spectral density satisfies (e.g., \citealp{brockwell-davis:2016a})
\begin{equation}
f(\lambda)=\frac{\sigma_\varepsilon^2}{2\pi}\,|C(e^{-i\lambda})|^2.
\label{eq:f_factorized_wold}
\end{equation}
Comparing \eqref{eq:f_factorized_basic} and \eqref{eq:f_factorized_wold} yields
\begin{equation}
|h(0)|^2=\frac{\sigma_\varepsilon^2}{2\pi},
\qquad\text{so that}\qquad
\sigma_\varepsilon^2 = 2\pi e^{a_0}.
\label{eq:sigmaeps}
\end{equation}

\subsection{From cepstral coefficients to Wold coefficients}

Under \eqref{eq:szego_condition}, the normalized spectral factor admits the
exponential representation
\begin{equation}
C(z)=\exp\!\left(\sum_{k=1}^{\infty} a_k z^k\right),
\qquad z\in\mathbb{D},
\label{eq:C_expansion}
\end{equation}
where $\{a_k\}_{k\ge 1}$ are the positive-index cepstral coefficients in
\eqref{eq:cepstral}. Expanding \eqref{eq:C_expansion} as a power series and
matching coefficients yields a recursion for $\{c_k\}$.
This can be written as
\begin{equation*}
c_{k+1} = \sum_{j=0}^{k}\left(1-\frac{j}{k+1}\right)a_{k+1-j}\,c_j,
\qquad k=0,1,2,\ldots,
\end{equation*}
initialized at $c_0=1$. This recursion is standard in the time-series
literature on spectral factorization and cepstral methods, see
\citet{pourahmadi:01a}. Related discussions of spectral factorization appear
in \citet{proietti-luati:19a} and \citet{bingham:2012a}.

\subsection{Numerical implementation}

In practice, we approximate the cepstral coefficients $a_k$ by evaluating $\log f(\lambda)$
on a fine equispaced midpoint grid (wrapped to $(-\pi,\pi]$) and computing the corresponding
discrete Fourier coefficients using an FFT. We then obtain $c_0,\ldots,c_q$ from the recursion
in \eqref{eq:pourahmadi_recursion} with $c_0=1$. The innovation variance $\sigma_\varepsilon^2$
is computed from \eqref{eq:sigmaeps}. The midpoint grid does not touch the singularity of $f$ at the
origin, so no further regularization is needed. Table \ref{tab:fgn_ck} reports the resulting coefficients for selected values of $H$.

\begin{table}[H]
\centering
\caption{Normalized Wold coefficients $c_k$ of fGn for various values of $H$.}
\label{tab:fgn_ck}
\scalebox{0.9}[0.9]{
\begin{tabular}{rrrrr}
\toprule
$k$ & $H=0.1$ & $H=0.2$ & $H=0.3$ & $H=0.4$ \\
\midrule
0 & 1.0000 & 1.0000 & 1.0000 & 1.0000 \\
1 & -0.6568 & -0.4588 & -0.2926 & -0.1419 \\
2 & -0.0622 & -0.0788 & -0.0711 & -0.0442 \\
3 & -0.0337 & -0.0443 & -0.0419 & -0.0273 \\
4 & -0.0224 & -0.0300 & -0.0292 & -0.0197 \\
5 & -0.0164 & -0.0224 & -0.0222 & -0.0153 \\
6 & -0.0128 & -0.0176 & -0.0178 & -0.0125 \\
7 & -0.0103 & -0.0144 & -0.0148 & -0.0105 \\
8 & -0.0086 & -0.0121 & -0.0126 & -0.0091 \\
9 & -0.0073 & -0.0104 & -0.0109 & -0.0080 \\
10 & -0.0063 & -0.0091 & -0.0096 & -0.0071 \\
\bottomrule
\end{tabular}}
\end{table}

\end{document}